\documentclass[acmsmall,nonacm]{acmart}

\renewcommand\footnotetextcopyrightpermission[1]{}

\usepackage[utf8]{inputenc}
\usepackage{enumitem}
\usepackage{xspace}
\usepackage[linesnumbered,vlined]{algorithm2e}
\usepackage[table]{xcolor}
\usepackage{wrapfig}
\usepackage{listings}
\usepackage{multirow}
\usepackage{pifont}
\usepackage{setspace}
\usepackage{hyperref}
\usepackage{pgfplots}
\pgfplotsset{compat=1.18}
\usepackage{fancyvrb}
\usepackage{lstlinebgrd}
\usepackage[most]{tcolorbox}
\usepackage[mathcal]{euscript}
\usepackage{makecell}

\graphicspath{{imgs/}}

\AtBeginDocument{%
  }

\definecolor{agentbg}{RGB}{255,235,230}
\definecolor{devbg}{RGB}{235,240,255}

\newtcolorbox{agentbox}{
  colback=agentbg,
  colframe=red!60!black,
  boxrule=0.5pt,
  arc=3pt,
  left=1pt,
  right=1pt,
  top=1pt,
  bottom=1pt,
  width=\linewidth,
  fonttitle=\bfseries,
  title=@Agent,
}

\newtcolorbox{devbox}{
  colback=devbg,
  colframe=blue!60!black,
  boxrule=0.5pt,
  arc=3pt,
  left=1pt,
  right=1pt,
  top=1pt,
  bottom=1pt,
  width=\linewidth,
  fonttitle=\bfseries,
  title=@Dev,
}
\newcommand{\cmark}[0]{\color{green!50!black}{\ding{51}}}

\newcommand{\ignore}[1]{}
\newcommand{\etc}{etc.\xspace}

\newcommand{\eg}{e.g.,\xspace}

\newcommand{\tool}{\textsc{WitGen}\xspace}

\newcommand{\myparagraph}[1]{\vspace{0.35em}\noindent\emph{#1.}}
\newcommand{\rn}[1]{\expandafter{\romannumeral #1\relax}}

\newcommand{\query}[0]{\textsc{Query}}
\newcommand{\dom}[1]{\mathrm{dom}(#1)}
\newcommand{\prog}[0]{\mathit{Prog}}
\newcommand{\frag}[0]{f_r}
\newcommand{\instr}[0]{i}
\newcommand{\M}[0]{\mathcal{M}}
\newcommand{\Mllm}[0]{\M^{\instr}_{\textsc{llm}}}
\newcommand{\Mdev}[0]{\M^{\instr}_{\textsc{dev}}}
\newcommand{\concat}{\mathbin{+\mkern-5mu+}}

\theoremstyle{definition}
\newtheorem{definition}{Definition}[section]
\newtheorem{example}{Example}[section]

\renewcommand{\equationautorefname}{\relax}
\def\equationautorefname#1{}

\definecolor{mycolor}{rgb}{0.122, 0.435, 0.698}
\definecolor{darkgreen}{RGB}{0,100,0}
\definecolor{darkred}{RGB}{100,0,0}
\definecolor{darkblue}{RGB}{0,0,150}
\definecolor{darkyellow}{RGB}{100,100,0}
\definecolor{darkmagenta}{RGB}{100,0,100}

\usepackage{tcolorbox}

\lstdefinestyle{CStyle}{
  language=C,
  basicstyle=\ttfamily\scriptsize,
  keywordstyle=\color{blue},
  morekeywords={string, iterator, list, char32_t},
  keywordstyle=[2]\color{teal},
  commentstyle=\color{gray}\itshape,
  stringstyle=\color{red},
  showstringspaces=false,
  numbers=none,
  escapeinside={(*@}{@*)},
  tabsize=2,
  xleftmargin=1em,
  frame=single,
  framesep=2pt,
  backgroundcolor=\color{gray!8},
}

\lstdefinestyle{CStyle}{
  language=C,
  basicstyle=\ttfamily\scriptsize,
  keywordstyle=\color{blue},
  morekeywords={ptrdiff_t,intptr_t},
  morekeywords=[2]{malloc,free},
  commentstyle=\color{gray}\itshape,
  stringstyle=\color{red},
  showstringspaces=false,
  numbers=none,
  escapeinside={(*@}{@*)},
  tabsize=2,
  xleftmargin=1em,
  frame=single,
  framesep=2pt,
  linebackgroundcolor={
    \color{black!5}
    \ifnum\value{lstnumber}>17
      \ifnum\value{lstnumber}<20
        \color{green!30}
      \fi
    \fi
    \ifnum\value{lstnumber}=20
      \color{red!25}
    \fi
  },
  literate={0}{{{\color{red}0}}}{1}
           {1}{{{\color{red}1}}}{1}
           {2}{{{\color{red}2}}}{1}
           {3}{{{\color{red}3}}}{1}
           {4}{{{\color{red}4}}}{1}
           {5}{{{\color{red}5}}}{1}
           {6}{{{\color{red}6}}}{1}
           {7}{{{\color{red}7}}}{1}
           {8}{{{\color{red}8}}}{1}
           {9}{{{\color{red}9}}}{1}
}

\lstdefinestyle{CStyleNumbers}{
  style=CStyle,
  numbers=left,
  numberstyle=\tiny\color{gray},
  stepnumber=1,
  numbersep=6pt
}

\begin{document}

\title{When is LLM-Based Program Reasoning Correct?}
\subtitle{A Completion Semantics for LLM-Based Code Inference}

\author{Zhiyuan Liu}
\affiliation{%
  \institution{Nanjing University}
  \city{Nanjing}
  \country{China}
}
\email{zhiyuanliu@smail.nju.edu.cn}

\author{Yihe Li}
\authornote{Joint first author}
\affiliation{%
  \institution{National University of Singapore}
  \city{Singapore}
  \country{Singapore}
}
\email{yihe.li@u.nus.edu}

\author{Trevor E. Carlson}
\affiliation{%
  \institution{National University of Singapore}
  \city{Singapore}
  \country{Singapore}
}
\email{tcarlson@nus.edu.sg}

\author{Huiyan Wang}
\affiliation{%
  \institution{Nanjing University}
  \city{Nanjing}
  \country{China}
}
\email{why@nju.edu.cn}

\author{Ruijie Meng}
\affiliation{%
  \institution{National University of Singapore}
  \city{Singapore}
  \country{Singapore}
}
\email{ruijie_meng@u.nus.edu}

\author{Gregory J. Duck}
\affiliation{%
  \institution{National University of Singapore}
  \city{Singapore}
  \country{Singapore}
}
\email{gregory@comp.nus.edu.sg}

\pagestyle{plain}

\begin{abstract}
Due to token and cognitive limits, {\em Large Language Models} (LLMs) typically perform program reasoning over {\em incomplete} code fragments/prompts rather than complete programs.
Such reasoning therefore must rely on {\em assumptions} about omitted code and context.
As a result, the meaning of an inference over a program fragment is not absolute, but depends on an implicit {\em completion model} describing how the fragment may be refined into a complete program.
In this paper, we introduce {\em completion semantics} for LLM-based program reasoning.
We formalize incomplete programs as denoting a space of possible refinements and define the correctness of existential inferences relative to a completion model.
Under this view, a reported bug is correct whenever there exists a completion within the model that witnesses the bug.
This perspective explains why many LLM-generated reports are neither simply correct nor incorrect, but instead depend on assumptions about omitted context.
We have instantiated our approach in the form of a witness-generation workflow that concretizes completions underlying an inference by constructing executable refinements of the original program fragment.
Witnesses serve both as evidence for existential claims and as a mechanism for exposing the assumptions required to support them.
We evaluate our approach on real-world LLM-generated bug reports and program-analysis tasks. Our results show that witness generation effectively distinguishes inferences supported by plausible completions from those requiring unrealistic assumptions, providing a practical mechanism for validating reasoning over incomplete programs.
\end{abstract}

\maketitle

\section{Introduction}\label{sec:intro}

{\em Large Language Models} (LLMs) are increasingly used for {\em program reasoning tasks}, such as code review~\cite{codereview, codereview2}, bug detection~\cite{AutoBug, meng2026hyllfuzz, chatafl}, security auditing~\cite{repoaudit, audit2}, and program comprehension~\cite{comprehension, comprehension2}.
In a typical workflow, an LLM is {\em prompted} with a task description (e.g., vulnerability detection, assertion violations, or safety conditions), together with a {\em code fragment} (e.g., a function, module, or program slice) for the LLM to reason over.
The LLM will proceed to reason over the fragment and instructions to produce an {\em outcome}, constructing an explanation, finding a bug, or determining that a property appears to hold.
Recent systems demonstrate how LLMs can reason effectively over code, and can identify behaviors that are difficult to capture using traditional program analyses.
As a result, LLM-based reasoning has become a central component of modern software engineering tools and coding agents, and forms the basis of many systems~\cite{AutoBug, wu2026palm, bouras2026defusing}.

Unlike traditional program analyses, LLMs rarely reason over {\em complete programs}, due to practical limitations---including token budgets, retrieval boundaries, and cognitive constraints~\cite{levy2024task, fang2024llm}. Rather, LLM-based program reasoning must typically operate on a {\em partial} view of a larger codebase.
As such, the prompt presented to the LLM typically only contains a {\em fragment} of the relevant program, such as an individual function together with a limited amount of surrounding context.
Consequently, any inference made by the LLM necessarily depends on assumptions about any code that has been excluded from the prompt.
In traditional program analysis, correctness is defined with respect to a complete program and its execution semantics.
In contrast, an incomplete code fragment, as part of an LLM prompt, does not denote a unique program.
The meaning of an inference over an incomplete fragment is therefore unclear:
{\em what does it mean for the LLM's conclusion to be correct?}

\begin{figure}
\begin{minipage}{0.39\linewidth}
\centering
\lstset{style=CStyle, numbers=left}
\begin{lstlisting}
 int f(struct S *s) {
   int *p = s->a;
   memset(s, 0, sizeof(*s));
   return p[0];  // Returns zero?
 }
\end{lstlisting}
\end{minipage}
\hfill
  \begin{minipage}[c]{0.57\textwidth}
    \caption{
    Example of an incomplete program fragment (e.g., a {\em prompt}).
    Here, the answer to the query ``{\em returns zero?}'' depends on assumptions made about the definition of \texttt{struct S}.
    } \label{fig:example}
  \end{minipage}
\end{figure}

\subsection{Illustrative Example}\label{sec:example-intro}

\autoref{fig:example} illustrates a simple example that highlights the problem with reasoning over incomplete program fragments.
Here, we consider a simple query: ``{\em does the function \texttt{f(...)} always return zero?}''
Classical program reasoning methods can attempt to answer this query using standard {\em points-to} ({\em alias}) and {\em data-flow} static analysis methods.
If \texttt{p} and \texttt{s} {\em alias} the same object in memory, then \texttt{f(...)} will indeed always return \texttt{0}; otherwise, a non-zero value may be returned.

However, the fragment in \autoref{fig:example} is {\em incomplete}.
Specifically, the definition of (\texttt{struct S}) has been omitted from the prompt, meaning that the answer to the query implicitly depends on {\em assumptions} about missing context.
For example, suppose that the \texttt{a} field is an {\em inline array field}, such as follows:
{\small
\begin{Verbatim}[commandchars=\\\{\}]
                                \textcolor{darkgreen}{struct} S \{ \textcolor{darkgreen}{int} a[\textcolor{darkred}{3}]; \};
\end{Verbatim}
}
In this case, \texttt{a} will be a sub-object of \texttt{*s}, meaning that the \texttt{memset(...)} operation will overwrite the storage referenced by \texttt{p}.
Under this assumption, the function {\em always returns zero}.
Alternatively, suppose that the \texttt{a} field is a {\em pointer field}, such as follows:
{\small
\begin{Verbatim}[commandchars=\\\{\}]
                                 \textcolor{darkgreen}{struct} S \{ \textcolor{darkgreen}{int} *a; \};
\end{Verbatim}
}
In this case, \texttt{memset(...)} will only overwrite the pointer itself, and leaves any memory referenced by \texttt{p} unchanged.
Under this assumption, the function {\em does not return zero}.
Thus, the fragment in \autoref{fig:example} admits {\em multiple completions}, and each yields a {\em different} answer to the same query.

This simple example highlights the fundamental challenge regarding reasoning over partial programs: the answer often depends on the omitted code and context.
For example, if we consider standard {\em data-flow} (\emph{def-use}) analysis over the fragment, the answer is context-dependent:

{\small
\[
\begin{array}{l|c|c|c}
\textbf{Definition of } \texttt{struct S}
&
\mathrm{Def}(\texttt{memset}(s,\ldots))
&
\mathrm{Use}(\texttt{p[0]})
&
\textbf{Returns zero?}
\\
\hline
\texttt{\textcolor{darkgreen}{int} a[\textcolor{darkred}{3}]}
&
\{\ell_S,\ell_S.a[0],\ell_S.a[1],\ell_S.a[2]\}
&
\{\ell_S.a[0]\}
&
\text{\em yes}
\\
\texttt{\textcolor{darkgreen}{int} *a}
&
\{\ell_S, \ell_S.a\}
&
\{\ell_H[0]\}
&
\text{\em not guaranteed}
\end{array}
\]
}%
Assuming \texttt{s->a} points to a location $\ell_H$ for the second case.
Modern LLMs routinely make such assumptions when reasoning over incomplete programs, yet the meaning of a conclusion relative to those assumptions remains largely undefined.
The central question of this paper is not whether an LLM's answer is correct in isolation, but {\em when} an LLM's answer should be considered correct.

\subsection{Completion Semantics}

The preceding example illustrates a more general observation: an incomplete program fragment does not denote a unique executable program, but rather a {\em space} of possible {\em completions}.
Here, a completion is the original fragment (unmodified) augmented with some additional code context to make a ``complete'' standalone program, including adding any missing functions, types, macros, globals, an entry-point, libraries, and environment.
Each completion therefore corresponds to a particular {\em interpretation} of the omitted code and context, and each completion may induce different answers to the same query (as illustrated in \autoref{sec:example-intro}).
Consequently, the correctness of a query may not be wholly defined with respect to any individual fragment alone.
Instead, correctness must be interpreted {\em relative} to a set of admissible completions.

In this paper, we formalize this intuition through {\em completion semantics}.
At a high level, a completion extends an incomplete program fragment into a self-contained program.
A {\em completion model} then characterizes which such completions are considered {\em admissible}.
Different completion models induce different notions of correctness.
For example, one model may admit all {\em syntactically valid completions}, while another may restrict completions that satisfy type constraints, API contracts, or other assumptions about intended program behavior.
Queries are subsequently interpreted relative to the given completion model.
Informally, an existential property (e.g., assertion failure) will hold if there exists an admissible element of the completion model witnessing the property, while a universal property holds if {\em all} admissible completions satisfy it.

Completion semantics are relevant for LLM-based program reasoning, since the boundary between provided and omitted context in prompts is generally unavoidable in practice.
While additional context can always be retrieved and appended to a prompt (e.g., the type definition in \autoref{sec:example-intro}), for real-world code, this usually shifts the boundary elsewhere in the program.
Modern software systems are sufficiently large that modern LLMs cannot effectively reason over an entire codebase and its environment simultaneously.
As a result, assumptions about omitted context are not an implementation artifact, but an inherent aspect of LLM-based reasoning itself.

A practical challenge is that completion models are often {\em implicit}: the assumptions underlying an LLM's reasoning are usually not directly observable.
To study these assumptions empirically, we instantiate our semantics using a {\em witness-generation} workflow that attempts to construct concrete completions supporting a reported inference.
Under our semantics, such witnesses serve as evidence for existential claims and expose the assumptions required for an inference to hold.
Witness generation is therefore one pragmatic realization of completions semantics.

In summary, our contributions are as follows:
\begin{itemize}[leftmargin=*]
\item We identify the problem of reasoning over incomplete programs and argue that the correctness of LLM-generated inferences should be defined relative to a {\em completion model};
\item We introduce {\em completion semantics}, a formal framework that interprets incomplete programs as spaces of possible refinements and defines query satisfaction relative to completion models;
\item We instantiate completion semantics through a {\em witness-generation workflow} for concretizing completions for code fragments in LLM prompts, and exposing assumptions underlying existential inferences; and
\item We evaluate the approach on real-world program-analysis tasks including bug reports, and demonstrate its utility for validating reasoning over incomplete code fragments.
\end{itemize}
Our completion semantics formalism is intentionally modest.
Our objective is not to introduce some deep new semantic machinery, but rather to identify the correct abstraction for reasoning over incomplete programs.
Our abstraction has surprising explanatory power: it accounts for a range of empirical phenomena previously studied in LLM-based program reasoning~\cite{nong2026llm, koterba2026llm, yang2025llm, li2025everything, fang2024llm}---including context sensitivity, disagreement between analyses, and several distinct classes of reasoning failures---while remaining independent of any particular reasoning engine or prompting strategy.
Accordingly, the witness-generation workflow should be viewed as an experimental vehicle for evaluating the semantics rather than as the primary contribution of the paper.

\section{Motivation}

In this section, we provide motivating examples, a problem statement, and summarize our approach.

\begin{figure*}[t]
\centering
\begin{minipage}{0.71\textwidth}
\begin{lstlisting}
CURLcode Curl_ws_request(struct Curl_easy *data, REQTYPE *req)
{
 unsigned char rand[16];
 char *randstr;
 size_t randlen;
 char keyval[40];
 struct SingleRequest *k = &data->req;

 /* ... */

 result = Curl_rand(data, rand, sizeof(rand));
 if(result)
  return result;
 result = (*@\texttt{Curl\_base64\_encode}@*)(rand, sizeof(rand), &randstr,&randlen);
 if(result)
  return result;
 DEBUGASSERT(randlen < sizeof(keyval));
 if(randlen >= sizeof(keyval))
  return CURLE_FAILED_INIT;
 strcpy(keyval, randstr);
 free(randstr);

 /* ... */

 k->(*@\texttt{upgr101}@*) = (*@\texttt{UPGR101\_WS}@*);
 return result;
}
\end{lstlisting}
\end{minipage}\hfill
\begin{minipage}{0.27\linewidth}
\footnotesize

\begin{agentbox}
Hello security team,\\
Hope you are doing well :)\\
I would like to report a potential security vulnerability in the
WebSocket handling code of the \texttt{curl} library. The issue is
related to the usage of the \texttt{strcpy} function, which can lead
to a buffer overflow if the length of the input is not properly checked.
The vulnerable code snippet is located at \emph{[line 20]}.

\emph{[Content abridged for brevity.]}
\end{agentbox}

\begin{devbox}
@Agent can you elaborate on (A) why the length check on \emph{[line 18]} is not
enough and (B) how the length can end up longer than the
\texttt{keyval} buffer?
\end{devbox}

\end{minipage}
\caption{Code fragment from \texttt{cURL} WebSocket handling reported as a buffer overflow vulnerability due to use of \texttt{strcpy}.
The correctness of the inference made by the LLM depends on assumptions made over the complete program.}
\label{fig:curl-strcpy}
\end{figure*}

\subsection{Example}\label{sec:example}

\autoref{fig:curl-strcpy} shows an 
AI-generated bug report submitted to the bug bounty program of the \texttt{cURL} project.\footnote{See \url{https://hackerone.com/reports/2298307} (retrieved July 2026).}
Here, the LLM flags a ``{\em potential security vulnerability in the WebSocket handling code of the curl library}'' related to the use of the \verb+strcpy+ function in the highlighted line (line 20).
The LLM has inferred that the call to \verb+strcpy+ ``{\em can lead to a buffer overflow if the length of the input is not properly checked}.''
Based on this inference, the report goes on to describe reproduction steps involving a ``{\em base64-encoded nonce value that exceeds the buffer size}'' and recommends replacing \verb+strcpy+ with \verb+strncpy+ as a safer alternative.
Although the original prompt was never made public, we can induce a similar bug report for line 20 using local LLMs (e.g., \texttt{llama3}) with the function and a prompt asking the existential query ``{\em does the following function have a buffer overflow?}''.
In this paper, our question is not whether LLMs can make mistakes (they can), but rather {\em what it means for a given LLM inference to be correct or not}?

The answer to this question depends on what {\em assumptions} can be made over the omitted code.
As illustrated in \autoref{fig:example} in \autoref{sec:example-intro}, a program fragment can be {\em completed} in different ways with different definitions, and the choice of the completion can change the result of the query---and by extension, the correctness of any inference made by the LLM.
For the bug report in \autoref{fig:curl-strcpy}, we consider two possible completions that result in different interpretations.

\subsubsection{Interpretation A: Contract-Preserving Completions}

The most natural interpretation of the example in \autoref{fig:curl-strcpy} is to assume that omitted code behaves according to its {\em intended contracts}.
Specifically, we can identify two main contracts over the return values of the \verb+curl_base64_decode+ call, specifically:
\begin{enumerate}[leftmargin=*]
\item \label{case:valid} \emph{Valid string}: The \texttt{randstr} string is a valid object and correctly null-terminated; and
\item \label{case:length} \emph{Length}: \texttt{randlen == strlen(randstr)}
\end{enumerate}
Under this interpretation, the safety check on lines 18--19 guarantees that the \verb+strcpy+ destination buffer on line 20 is sufficiently large for the copied string.
Consequently, no buffer overflow can occur, meaning that the original LLM inference is \textbf{incorrect}.

The actual \texttt{cURL} implementation satisfies this interpretation, but the argument does not depend on the concrete codebase itself in general.
Rather, it relies on the broader assumption that omitted functions respect their intended behavior and maintain consistent length and string invariants.
For example, contract (\ref{case:valid}) is a common assumption for all strings passed on over function boundaries in \verb+C+ programs.
Contract (\ref{case:length}) can be inferred by a natural language interpretation over the $(\texttt{randstr}, \texttt{randlen})$ variable pair.
Specifically, that \texttt{randstr} returns a string, the common \texttt{rand}-prefix shared between variables, and the \texttt{len}-suffix are all heuristic clues that \texttt{randlen} is intended to return the length of \texttt{randstr}.
Both humans and LLMs alike will make such assumptions.

\subsubsection{Interpretation B: Contract-Agnostic Completions}\label{sec:interpret-B}
Instead of relying on heuristics, we could allow completions that ignore implied contracts.
Under this relaxed model, we consider any completion to be valid provided it is {\em syntactically correct} and does not redefine any symbol.
However, under this model, it turns out that the {\em call to \verb+strcpy+ on line 20 can actually overflow}, despite the safety check on lines 18--19.
Under this interpretation, the original LLM inference must be deemed {\bf correct}.

To see why, we consider a completion of \verb+Curl_base64_encode+ that breaks the implied contracts (\ref{case:valid}) and (\ref{case:length}).
For example, we can consider a completion with the following function definition:

{\small
\begin{Verbatim}[commandchars=\\\{\}, frame=single]
  CURLcode \textcolor{darkblue}{curl_base64_decode}(\textcolor{darkgreen}{const char} *, \textcolor{darkgreen}{size_t}, \textcolor{darkgreen}{char} **outstr, \textcolor{darkgreen}{size_t} *outlen) \{
    *outptr = \textcolor{darkblue}{malloc}(\textcolor{red}{40});
    \textcolor{darkblue}{memset}(*outptr, \textcolor{red}{'A'}, \textcolor{red}{40});      \textcolor{darkmagenta}{// Violation since *outlen is not nul-terminated}
    *outlen = \textcolor{red}{20};                  \textcolor{darkmagenta}{// Violation since *outlen != strlen(*outstr)}
    \textcolor{darkyellow}{return} CURL_OK;
  \}
\end{Verbatim}
}
This function is syntactically correct and compiles without error.
If this definition were allowable, then the overflow on line 20 can be induced as follows:
\begin{enumerate}[leftmargin=*]
\item The call to \verb+Curl_base64_encode+ returns a non-null-terminated string of length 40;
\item The safety check on line 18 passes since \verb+randlen+ is 20 and \verb+sizeof(keyval)+ is 40; and
\item The \verb+strcpy+ operation on line 20 {\em overflows} given the non-null terminated \verb+randstr+.
\end{enumerate}
This example further highlights how inference over incomplete program fragments implicitly depends on assumptions about the missing code, and how even questionable LLM inference could actually be valid under some interpretations.
The question arises, when should we consider a completion to be valid, and when it is not?

\subsection{Problem Statement}\label{sec:problem}

The example in \autoref{fig:curl-strcpy} demonstrates that correctness cannot be defined relative to arbitrary completions alone.
If every syntactically valid completion is admitted, then many properties become trivially realizable by constructing sufficiently pathological implementations of omitted code.
In the present example, the reported overflow can be induced by a completion that violates natural assumptions regarding string termination, length consistency, and API behavior.
Under such a model, the bug report must be considered correct, despite relying on assumptions that most developers would regard as unreasonable.

At the opposite extreme, we may attempt to restrict completions using heuristically inferred contracts and domain knowledge.
For example, \verb+C+ strings are overwhelmingly null-terminated when passed across function boundaries, and paired variables such as $(\texttt{randstr}, \texttt{randlen})$ strongly suggest a length-string relationship.
Both humans and LLMs routinely rely on such assumptions when reasoning about code.
Under these assumptions, the report in \autoref{fig:curl-strcpy} is incorrect.
However, such assumptions are also not universally valid.
Familiar library interfaces such as \texttt{strncpy}, \texttt{strxfrm}, and \texttt{readlink} violate common string-handling conventions, while real-world software frequently contains undocumented behavior, incomplete specifications, and implementation-specific invariants.
Consequently, correctness cannot be reduced to heuristic assumptions alone.

The challenge is therefore to characterize which completions should be considered {\em admissible} when reasoning over an incomplete program fragment.
On the one hand, purely syntactic notions of completion are too permissive and admit unrealistic interpretations.
On the other hand, informal notions of intended behavior are difficult to define precisely and may vary across developers, domains, and programming environments.
The problem is further complicated by the fact that LLMs themselves are often treated as {\em black boxes}: while an inference may implicitly depend on assumptions about omitted context, those assumptions are generally not made explicit.

This paper takes the position that reasoning over incomplete programs should be understood in terms of a {\em completion model}: a set of admissible completions against which queries are interpreted.
Rather than attempting to define correctness directly in terms of LLM outputs, we instead define correctness relative to a completion model and treat the LLM as a mechanism for exploring that model.
This perspective separates the semantic question---{\em what completions are considered valid?}---from the algorithmic question of how a particular model, human, or analysis procedure reasons about those completions.

\subsection{Our Approach}\label{sec:approach}

We interpret an incomplete program fragment as denoting a space of possible completions rather than a single program.
A completion model specifies which of these completions are admissible, thereby determining the assumptions under which omitted code is interpreted.
Program properties are then evaluated relative to the completion model: existential queries hold whenever some admissible completion witnesses the property, whereas universal queries must hold for every admissible completion.

\begin{figure}[t]
    \centering
    \includegraphics[width=.95\linewidth]{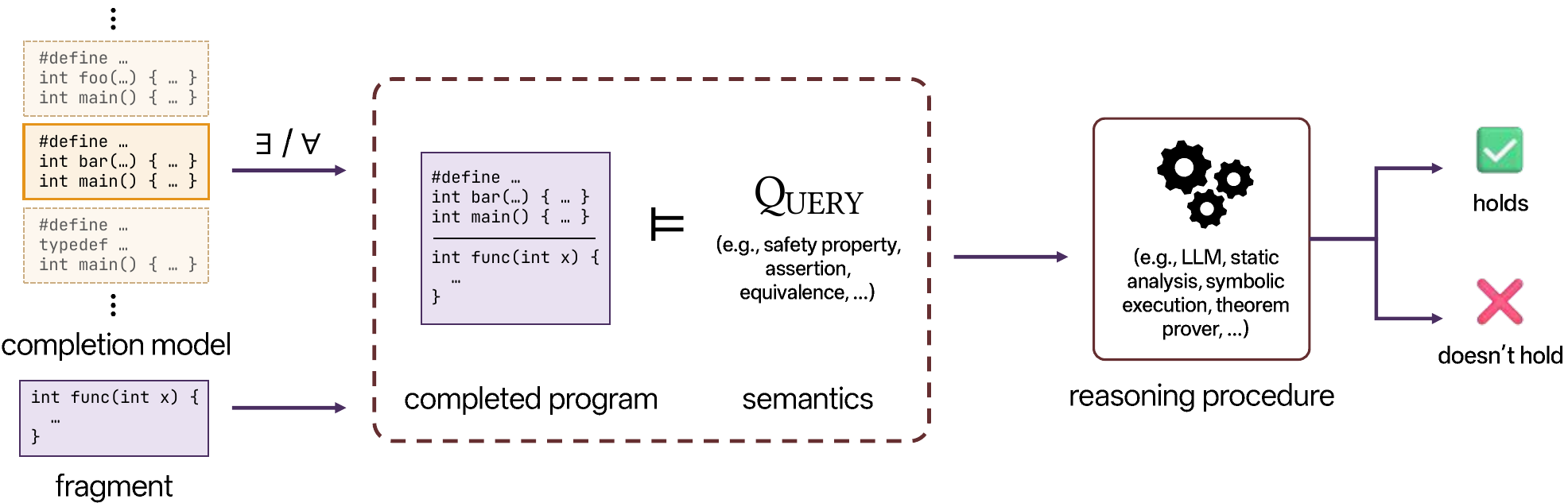}
    \caption{Illustration of completion semantics for reasoning over incomplete programs.}
    \label{fig:completion-sets}
\end{figure}

A key feature of our approach is that we {\em separate} three different concerns that can be conflated when using LLMs for program reasoning:
\begin{enumerate}[leftmargin=*]
\item Firstly, the underlying {\em program semantics} determines {\em when} a property should hold, but generally only over a complete program;
\item Secondly, the {\em completion model} determines {\em how} program fragments should be interpreted, and which assumptions about omitted context are considered admissible; and
\item Finally, a {\em reasoning procedure} determines {\em whether} properties can be established within those completions.
\end{enumerate}
This separation allows reasoning over incomplete programs to be studied independently of any particular analysis technique or language model:
a program {\em fragment} implies a {\em completion model}, which defines the {\em semantics}, which can be decided via a {\em reasoning procedure}.
Our approach is illustrated in \autoref{fig:completion-sets}.

Our framework is designed for LLM-based program reasoning.
When presented with an incomplete code fragment, an LLM must implicitly make assumptions about context that is not present in the prompt.
These assumptions may concern missing implementations, programmer intent, coding conventions, undocumented contracts, or other latent information.
Completion models provide a formal abstraction of such assumptions.
Under this view, an LLM-generated inference may be understood as a claim about the existence or absence of completions supporting the reported property.
Failure of this chain is not as simple as ``{\em hallucination}'', but could arise from query defects/misinterpretation, the completion model being too restricted or permissive, or a failure of the reasoning procedure itself.

A final practical challenge is that completion models are often {\em implicit}.
For example, LLM-based reasoning is often treated as a {\em black-box}, meaning that the underlying assumptions that an LLM uses for inference are typically not directly observable.
However, it may be possible to use the same LLM to generate a completion that {\em supports} (or {\em refutes}) a specific inference, and to also concretize any assumption used by the LLM in the inference.
This is a form of {\em witness generation}:
rather than treating the LLM as an opaque oracle, we generate a concrete completion (a witness) to {\em expose} the assumptions necessary for an inference to hold over the original fragment.

\section{Completion Semantics}\label{sec:semantics}

In this section, we formalize {\em completions} and {\em completion models} for reasoning over incomplete program fragments.

\subsection{Fragments and Completions}

An incomplete program fragment does not denote a single executable program, but rather admits many possible {\em completions}---each corresponding to a different interpretation of the omitted context.
For example, the fragment in \autoref{fig:example} may be completed with different definitions of \texttt{struct S}, leading to different answers to the same query.

To capture this intuition, we model both {\em fragments} and {\em completions} as {\em strings} of program text.

\begin{definition}[Completion]\label{def:completion}
Let $\frag$ be a {\em program fragment} represented as a {\em string}.
A {\em completion} of $\frag$ is any string $C$ that extends $\frag$ by supplying additional program context while preserving the contents of $\frag$.
We write $C \sqsupseteq \frag$ to denote that $C$ is a completion of $\frag$.
\end{definition}

\noindent
Our definition is intentionally permissive.
A completion need not compile, execute, or satisfy any semantic property.
It needs only to extend the original fragment.
More restrictive notions of completion, such as {\em syntactically} valid programs, {\em compilable} programs, or {\em contract-respecting} programs, will be introduced formally as {\em completion models}.

Furthermore, a precise definition of the $\sqsupseteq$ operation is also intentionally left open.
One definition could be the {\em substring} relation, i.e., $C = \mathit{prefix} \concat \frag \concat \mathit{postfix}$, which is our assumed default.
In general, more complicated definitions are possible.

\subsection{Completion Models}\label{sec:model-def}
The unrestricted notion of a completion is intentionally permissive.
In practice, not all completions are equally meaningful.
For example, since a completion can be an arbitrary string, almost all completions will fail to parse, will violate language semantics, or will rely on assumptions that would be considered unreasonable by a developer (see \autoref{sec:interpret-B}).
To address this issue, we introduce the notion of a {\em completion model}.
A completion model characterizes which completions are considered {\em admissible} when reasoning about an incomplete program fragment.

\begin{definition}[Completion Model]
Let $\frag$ be a program fragment. A {\em completion model} $\M(\frag)$ is any set of completions of $\frag$.
\end{definition}

\noindent
A completion model may be viewed as a filter over the unrestricted completion space. Different models admit different completions and therefore induce different interpretations of the same fragment. The examples in \autoref{sec:example} illustrate this phenomenon: a reported bug may be realizable under one completion model (e.g., all {\em compilable} completions) while impossible under another (assumed {\em developer} contracts).

The simplest completion model is the {\em unrestricted model}, which admits every completion of a fragment:
$\M_{\textsc{all}}(\frag) = \{C \mid C \sqsupseteq \frag \}$.
This model is usually far too permissive, as it admits completions containing arbitrary syntax errors, inconsistent type definitions, or implementations that deliberately violate the intent of the surrounding code.
More restrictive models are constructed by imposing additional constraints on admissible completions.
For example:
\begin{itemize}[leftmargin=*]
\item[-] {\em Syntactic models} only admit completions that parse successfully under the language grammar;
\item[-] {\em Compilation models} admit only completions that satisfy the language's static semantics (e.g., type checking, linking, and having an entry-point);
\item[-] {\em Contract models} that restrict completions according to programmer-supplied specifications, inferred contracts, API documentation, or other behavioral assumptions; and
\item[-] {\em Domain-specific models} that incorporate any other additional assumption about a particular programming environment, framework, or application domain.
\end{itemize}
These examples illustrate an important property of completion semantics: correctness is always relative to a completion model.
A query that is true under one model may be false under another (see \autoref{sec:example}).
The choice of completion model therefore determines which assumptions about omitted context are considered admissible.

In the context of LLM-based program reasoning, the most interesting completion model is the one implicitly assumed by the model itself.
Intuitively, an LLM does not reason over all possible completions of a fragment, but rather over some subset that it considers {\em plausible} given the prompt, training data, and inferred intent.
While such a completion model is not directly observable, it provides a useful abstraction for understanding the assumptions underlying LLM-generated inferences.
We return to this idea in \autoref{sec:llm-inference}.

\subsection{Query Satisfaction}

The preceding sections define how incomplete program fragments induce spaces of possible completions. We now define how queries are interpreted over such spaces.

Completion semantics does not replace existing notions of program semantics.
Rather, it builds upon them. We assume some underlying (possibly partial) semantic relation:
\begin{align}
\prog \models \query
\tag{\sc Semantic-Relation}
\end{align}
between complete programs $\prog$ and properties $\query$.
The precise interpretation of $\models$ and the $\query$ language depends on the analysis/semantics being applied.
For example, $\query$ may denote an {\em execution} property (e.g., assertion failure or memory safety), and a {\em static} property (e.g., type correctness).
Furthermore, the relation is typically only defined for
completions within the {\em semantic domain} of the query, written $\dom{\query}$.
For example, if $\query$ is an execution property,
then $\dom{\query}$ will typically contain all {\em compilable} and {\em runnable} programs.
Likewise, if $\query$ is a type-correctness property, then $\dom{\query}$ will typically contain all {\em parseable} programs.
The $\query$ itself can be expressed in any language of choice, including formal languages, and even natural language (common for LLM prompting).

The role of completion semantics is to {\em lift} the semantic relation from complete programs to incomplete program fragments.
Given a fragment $\frag$ and a completion model $\M(\frag)$, queries are interpreted relative to the completions admitted by the model:

\begin{definition}[Existential Satisfaction]\label{def:exists}
A completion model $\M(\frag)$ {\em existentially} satisfies a $\query$ if {\em there exists} a completion $C \in \M(\frag)$ such that $C \models \query$, i.e.:
\begin{align}
\M(\frag) \models_\exists \query
 \qquad \text{iff} \qquad
\exists C \in \M(\frag) \cap \mathrm{dom}(\query) : C \models \query
\tag{\sc Existential}
\end{align}
\end{definition}
\noindent
Intuitively, an existential query holds whenever at least one admissible completion witnesses the property. Many program-analysis tasks naturally take this form.
For example, a {\em bug report} can be interpreted as the claim that there exists a completion under which an assertion can fail or a vulnerability can be triggered under the program's execution semantics.

Likewise, we can define a {\em universal query}:
\begin{definition}[Universal Satisfaction]\label{def:forall}
A completion model $\M(\frag)$ {\em universally} satisfies a $\query$ if {\em for all} completion $C \in \M(\frag)$ we have that $C \models q$, i.e.:
\begin{align}
\M(\frag) \models_\forall \query
 \qquad \text{iff} \qquad
\forall C \in \M(\frag) \cap \mathrm{dom}(\query): C \models \query
\tag{\sc Universal}
\end{align}
\end{definition}
\noindent
Universal queries correspond to properties that must hold regardless of how the omitted context is completed.
For example, for \autoref{fig:example}, the query ``{\em does \texttt{f(...)} always return zero?}'' is an example of a universal property.
However, the answer depends on the completion model: if the model admits both pointer-based and array-based definitions of \texttt{struct S}, then the property does {\em not} hold universally.
The distinction gives rise to three useful notions:
\begin{itemize}[leftmargin=*]
\item[-] A query is {\em realizable} if it holds for at least one completion: $\M(\frag) \models_\exists \query$
\item[-] A query is {\em locally true} if it holds for every completion: $\M(\frag) \models_\forall \query$
\item[-] A query is {\em locally false} if it holds for no completion: $\M(\frag) \not\models_\exists \query$
\end{itemize}
An example of a {\em locally true} or {\em false} property could be ``{\em is the function correctly formatted?}'' under a supplied coding standard represented as ($\models$).
Such a query depends only on the fragment $\frag$, and is independent from any context.

\subsection{Discussion}

Completion semantics provides a very general mechanism for {\em lifting} existing reasoning techniques from complete programs to incomplete program fragments.
Any analysis capable of evaluating a property over a complete program may be used as the underlying semantic relation.
Examples include {\em execution-based} techniques such as {\em testing}, {\em symbolic execution}, and {\em model checking}, as well as static analyses such as {\em type checking}, {\em data-flow analysis}, and {\em abstract interpretation}, or even syntactic properties such as {\em code formatting standards}.
Completion semantics remains agnostic to the particular analysis being employed; its role is simply to interpret the resulting property relative to a completion model.

Viewed through this lens, the examples from \autoref{fig:example} and \autoref{fig:curl-strcpy} illustrate two distinct sources of ambiguity that arise when reasoning about incomplete programs.
In \autoref{fig:example}, the truth of the query depends on omitted type information, and different completions of \texttt{struct S} induce different program behaviors, leading to different answers to the same question.
In \autoref{fig:curl-strcpy}, the ambiguity is more subtle.
While a vulnerability may be realizable under some completions, many such completions rely on assumptions that violate the apparent {\em contracts} and {\em conventions} of the surrounding code.
The challenge is therefore not merely to determine whether a witness exists, but to determine which completions should be considered {\em admissible}.

These examples illustrate an important separation of concerns.
Traditional program semantics determines whether {\em a property holds for a complete program}.
Completion models determine which assumptions about omitted context are considered {\em admissible}.
Reasoning procedures determine how properties are established within those completions.
Completion semantics isolates these concerns and allows them to be studied independently.
As a consequence, disagreements about the validity of an inference can often be traced to differences in completion models rather than differences in the underlying program semantics.

\begin{example}[\texttt{cURL} Bug Report Revisited]
The distinction can be illustrated using the \texttt{cURL} example from \autoref{fig:curl-strcpy}.
Here, the underlying {\em semantics} is simply the execution semantics of the completed program: a bug exists if an execution reaches the reported overflow.
The {\em completion model} determines which implementations of the omitted context are considered admissible.
For example, a reasonable completion model may only contain definitions for \verb+Curl_base64_encode(...)+ that produce a nul-terminated string whose reported length is consistent with the generated output.
Finally, the {\em reasoning procedure} determines how the property is established.
One analysis may search for a violating execution using testing or fuzzing, while another may use symbolic execution to prove that no such execution exists.
In fact, under this completion model and symbolic execution, all paths to the \texttt{strcpy} operation must satisfy:
\begin{align}
\mathit{strlen}(\texttt{randstr}) < \mathit{sizeof}(\texttt{keyval}) \tag{\sc Path-Constraint}
\end{align}
 meaning that no \texttt{strcpy} overflow is possible.
These choices are independent: the same completion model may be analyzed using different reasoning procedures, while the same reasoning procedure may yield different conclusions under different completion models.
\end{example}


\section{Completion Models for LLM-Based Program Reasoning}\label{sec:llm-inference}

Applying completion semantics to LLMs is straightforward: the prompt defines an incomplete fragment, and the LLM implicitly reasons relative to an (unknown) completion model.

\subsection{LLM Completion Models}\label{sec:model}

Completion semantics can be applied to any setting in which program reasoning is performed over incomplete program fragments.
LLM-based program reasoning is a particularly important instance of this problem because the LLM is rarely prompted with a complete program (due to token limits and cognitive resources).
Instead, the prompt necessarily contains only a fragment of a larger codebase together with a task description and a limited surrounding context.
Consequently, any inference produced by the model necessarily depends on assumptions regarding code that is not present in the prompt fragment.
This observation suggests a natural interpretation of LLM-based reasoning in terms of completion models.
Given a {\em prompt}, comprising a program fragment $\frag$ and instruction $\instr$, we associate the LLM with an {\em implicit completion model} $\Mllm(\frag)$.
This model represents the set of completions that the LLM considers {\em plausible} when reasoning over fragment $\frag$ under instruction $\instr$.

\subsubsection*{Completion Model Definition}
Unlike some of the example completion models introduced in \autoref{sec:semantics}, $\Mllm(\frag)$ will generally not have a self-contained definition.
For example, a compiler-defined completion model can be characterized by the language specification.
Likewise, a contract-based completion model can be characterized by a formal description of explicit behavioral constraints.
In contrast, the assumptions underlying an LLM's reasoning emerge from training data, instruction ($\instr$) interpretation, and inferred context.
As a result, the corresponding completion model may generally lack an explicit descriptive definition.
That said, for our purposes, completion semantics does not require $\Mllm(\frag)$ to have an explicit, self-contained, definition.
Rather, the purpose of the model is not to describe the internal operation of the LLM, but rather to provide a semantic interpretation of the inferences it produces.
In particular, the completion model serves as an abstraction of the assumptions that must hold for a reported conclusion to be valid.

\subsubsection*{Completion Model Membership}
Even though $\Mllm(\frag)$ may generally lack a descriptive definition, it is nevertheless possible to construct individual completions that are consistent with the model's assumptions.
This can be achieved by constructing candidate completions and instructing the LLM to {\em test} for validity ($\Mllm(\frag)$ membership) under the LLM's internal assumptions.
Alternatively, we may also instruct the LLM to {\em generate} candidate completions.
This observation will form the basis of self-validation via witness generation that we shall explore further in \autoref{sec:witness}.

Unlike compiler- or contract-defined completion models, there is no guarantee that elements of $\Mllm(\frag)$ correspond to valid or executable programs.
An LLM may reason using assumptions that are incomplete, inconsistent, or otherwise unrealizable.
Likewise, the LLM may also make mistakes when testing for, or generating, element membership of the completion set.
This is a possible failure mode, and we shall elaborate on this issue later in \autoref{sec:failure}.

\subsection{Reasoning over Completion Models}\label{sec:reasoning}

The completion model $\Mllm(\frag)$ provides a semantic interpretation of the assumptions underlying an LLM's reasoning.
The remaining question is how inferences produced by the model should be interpreted relative to this completion model.
Under completion semantics, an LLM-generated inference is viewed as a claim regarding the satisfaction of a query over $\Mllm(\frag)$.
For example, suppose the model reports that a program fragment contains a vulnerability (i.e., is {\em unsafe}).
Such a report is interpreted as an {\em existential} claim stating that there exists a completion under which the vulnerability is realizable.
This is represented as follows:
\begin{align}
\Mllm(\frag) \models_{\exists} \textsc{Bug}
\tag{\sc Existential-Reasoning}
\end{align}
Likewise, a report asserting that a fragment is {\em safe} is interpreted as a universal claim stating that this property holds for all admissible completions:
\begin{align}
\Mllm(\frag) \models_{\forall} \textsc{Safe}
\tag{\sc Universal-Reasoning}
\end{align}
More generally, completion semantics do not prescribe which lifted satisfaction relation should be used for a given query.
Rather, the interpretation depends on the intended meaning of the instruction $\instr$ from the prompt, which is typically expressed in natural language.
For example, bug-finding instructions are naturally interpreted as existential queries, whereas safety or verification instructions are naturally interpreted as universal queries.
Completion semantics provides a formal meaning once the interpretation is fixed.

Our perspective also highlights the key distinction between traditional program reasoning and reasoning over incomplete programs.
Given a complete program, the primary question is whether a property holds under the underlying program semantics.
In contrast, reasoning over an incomplete fragment requires an additional layer of interpretation concerning the omitted context.
Consequently, the validity of an LLM-generated inference depends not only on the correctness of the reasoning process itself, but also on the assumptions embodied by the completion model.

\subsubsection*{Examples}
Returning to the examples from \autoref{fig:example} and \autoref{fig:curl-strcpy}, the reported conclusions can now be interpreted as claims over completion models.
In \autoref{fig:example}, the query is ``{\em does the function \texttt{f(...)} always return zero?}''.
The answer to the query depends on whether the completion model admits pointer-based definitions of \texttt{struct S}.
Most modern LLMs consider completions with the pointer-based definition to be {\em admissible}.
This means that the correct inference for the query must be ``{\em false}'', ``{\em not guaranteed}'', or equivalent.

Likewise, for \autoref{fig:curl-strcpy}, the reported vulnerability depends on whether the completion model admits implementations of \verb+Curl_base64_encode(...)+ that violate the apparent contracts of the surrounding code.
As shown in \autoref{sec:example}, if we allow for syntactic and compilable completions, then the inference of the vulnerability is {\bf correct}.
However, such models tend to be too permissive.
Instead, we can test for self-consistency against the LLM's own completion model.
To do so, we instruct the LLM to test candidate completions, such as the one from \autoref{sec:interpret-B}, for feasibility.
Assuming GPT-5, the LLM explicitly rejects this completion for violating implied contracts.
Specifically, the proposed completion:
\begin{quote}
``{\em is inconsistent with how a string-returning base64 routine is ordinarily expected to behave.}''
\end{quote}
Similarly, if we ask for GPT-5 to {\em generate} a completion for a reported vulnerability, it reports:
\begin{quote}
``{\em I do {\bf not} see a reasonable implementation of the base64 routine that both (1) satisfies the apparent API contract, and (2) causes the \verb|strcpy()| into \verb|keyval[40]| to overflow.}'' (emphasis original)
\end{quote}
Assuming these answers accurately reflect the completion model $\Mllm(\frag)$ under GPT-5 and this fragment, then any inference that \autoref{fig:curl-strcpy} contains a real vulnerability must be {\bf incorrect}.

\subsubsection*{Discussion}

Completion semantics provides a meaning for LLM-generated inferences independently of how those inferences are produced.
Whether the underlying reasoning procedure is a human developer, a language model, a symbolic executor, or some combination thereof, the resulting conclusion can be interpreted relative to a completion model.
This provides a common semantic framework for reasoning about incomplete programs and the assumptions required to make such reasoning meaningful.

We are also careful to point out that completion semantics themselves are a model for explaining inference, and not how LLMs actually work internally.
For example, completion semantics can explain {\em failure modes} as discussed next.

\subsection{Failure Modes}\label{sec:failure}

A useful consequence of completion semantics is that it makes explicit the distinct ways in which LLM-based program reasoning can fail.
Such failures are often grouped together under broad terms such as ``{\em hallucination}''.
However, once reasoning is interpreted relative to a completion model, it becomes possible to distinguish failures arising from incorrect assumptions about omitted context from failures arising from incorrect reasoning over those assumptions.
We explore these ideas in this subsection.

\subsubsection{Completion Model Errors (Prompting Failure)}
The first class of failures arises when the completion model itself is {\em misaligned} with the intended interpretation of the fragment or prompt instructions.
Informally, we may view the developer as inducing an idealized completion model $\Mdev(\frag)$ that captures the knowledge and assumptions intended for the surrounding code, and $\Mllm(\frag)$ is often an {\em abductive} approximation of the developer's intent guided by training.
However, there is no guarantee that $\Mllm(\frag)$ will actually coincide with $\Mdev(\frag)$ for any given code fragment.
As such, an LLM may admit or include completions that violate implicit contracts or other assumptions.
For example, consider the simple code fragment:

{\small
\begin{Verbatim}[commandchars=\\\{\}, frame=single]
void \textcolor{darkblue}{my_copy}(\textcolor{darkgreen}{char} *dst, \textcolor{darkgreen}{const char} *src, \textcolor{darkgreen}{size_t} n) \{
    \textcolor{darkyellow}{for} (\textcolor{darkgreen}{size_t} i = \textcolor{darkred}{0}; i < n; i++) dst[i] = src[i];
\}
\end{Verbatim}
}
Without additional context, an LLM may abductively assume a \verb+memcpy+-style contract for this function (non-\verb+NULL+ arguments for $\mathtt{n} > 0$, buffers valid for $\mathtt{n}$ bytes, no buffer overlap), and the completion model $\Mllm(\frag)$ will reflect this.
However, it is possible that the developer intended a different implied contract (e.g., overlapping buffers allowed, or \verb+dst+ being \verb+NULL+ is allowed as a {\em nop}), meaning that the code is actually buggy.
In such cases, the resulting inference may be technically correct with respect to $\Mllm(\frag)$ while nevertheless being unhelpful in practice.

Likewise, LLM-based inference may also fail due to misinterpretation of the instruction ($\instr$).
Completion semantics defines both an {\em existential} ($\models_\exists$) and a {\em universal} ($\models_\forall$) definition of satisfaction, and the correct definition to apply is usually implicit in the instruction.
For example, ``{\em does this code contain a bug?}'' implies existential reasoning, whereas ``{\em does this loop always terminate?}'' implies universal reasoning.
In principle, an LLM could misunderstand instructions and apply the wrong reasoning.
Such errors can be broadly classified as {\em prompting} or {\em interpretation} errors, and indicate that the prompt was insufficient to constrain the LLM's interpretation.
This is an inherent failure mode of LLM-based reasoning in general.

\subsubsection{Reasoning Errors (Hallucinations)}
The second class of failures arises when the completion model is reasonable, aligned with developer expectations, and the instruction is interpreted correctly; however, the reasoning process itself is incorrect.
In this case, the LLM fails to correctly determine whether or not a property holds relative to its own completion model.
Since LLMs are fundamentally approximate reasoning engines, they are prone to making mistakes.
For example, suppose that:
\begin{align*}
\Mllm(\frag) \models_{\exists} \textsc{Bug}
\end{align*}
holds, since there is a concrete element $C \in \Mllm(\frag)$ which witnesses the $\textsc{Bug}$ (i.e., the LLM agrees that $C$ is a valid completion), and $C \in \mathrm{dom}(\textsc{Bug})$.
However, the LLM may still erroneously report that no such bug exists.
Likewise, the converse may occur, where the model reports a bug exists, despite no admissible completion witnessing the property.

Such failures correspond most closely to the classical notion of a {\em hallucination}: the interpretation and assumptions used by the LLM are reasonable and aligned, but the inference itself is incorrect.

\subsubsection{Recovery Errors}
A third class of failures arises when attempting to recover or reconstruct the LLM's reasoning, e.g., by finding elements of the completion model.
As discussed in \autoref{sec:model}, LLM completion models generally lack an explicit descriptive definition.
Consequently, practical techniques, such as membership testing and completion generation, may themselves be imperfect.

For example, the LLM may incorrectly reject a completion that is consistent with its own assumptions, incorrectly accept an inconsistent completion, or generate a completion that does not lie within the semantic domain of the query (i.e., $C \not\in \textrm{dom}(q)$).
Likewise, external validation procedures may fail to recognize a valid witness due to incompleteness of the underlying oracle, resource limits, or timeouts. These failures do not necessarily imply that the completion model or reasoning process is incorrect per se; rather, these arise from the difficulty of recovering explicit elements from an implicit completion model.

\section{Witness Generation Instantiation}\label{sec:witness}

Completion semantics provides a general interpretation of LLM-generated inferences relative to completion models.
While the completion model induced by an LLM may lack an explicit descriptive definition, we can nevertheless use the LLM to {\em self-validate} under its own completion model using a {\em witness generation} workflow.
By instantiating completion semantics with an execution-based oracle, witness generation provides a method to {\em test} whether LLM-based program reasoning is correct for certain classes of queries.
Witness generation is not part of completion semantics itself.
Rather, it is one practical instantiation that makes completion models observable and allows the proposed semantics to be evaluated empirically.

\subsection{Completion Witnesses}\label{sec:witness-def}

Although we interpret inference as a claim regarding the existence of a completion satisfying some property, the completion itself is typically not directly observable.
To address this problem, we introduce the notion of a {\em completion witness}.
A completion witness is a concrete completion that justifies an {\em existential inference}, defined as follows:

\begin{definition}[Completion Witness]\label{def:comp-witness}
Let $\frag$ be a fragment, $\query$ a query, and $\M(\frag)$ a completion model.
Then $W$ is a {\em completion witness} for the existential query
$\M(\frag) \models_\exists \query$ iff:
(1) $W \in \M(\frag)$, (2) $W \in \dom{\query}$, and 
(3) $W \models \query$.
\end{definition}
\noindent
In other words, a {\em completion witness} is a completion that both satisfies the assumptions of the completion model, is within the semantic domain of the query, and demonstrates the queried property under the underlying semantics, satisfying \autoref{def:exists}. 
Completion witnesses are evidence for {\em existential} queries:
\begin{itemize}[leftmargin=*]
\item[-] The existence of a completion witness for $\query$ establishes $\M(\frag) \models_{\exists} \query$; but
\item[-] The failure to construct such a witness does {\em not} imply
$\M(\frag) \not\models_{\exists} \query$.
\end{itemize}
Likewise, a completion witness for the negated query can be used to {\em refute} universal queries:
\begin{itemize}[leftmargin=*]
\item[-] The existence of a completion witness for $\neg \query$ establishes $\M(\frag) \not\models_{\forall} \query$; but
\item[-] The failure to construct such a witness does {\em not} imply
$\M(\frag) \models_{\forall} \query$.
\end{itemize}
Consequently, completion witness generation can be viewed as a practical validation (or refutation) mechanism rather than a decision procedure.

\subsection{Completion Witness Generation for Self-Validation}

Suppose an LLM reports an existential inference:
\begin{align*}
\Mllm(\frag) \models_{\exists} \textsc{Query}.
\end{align*}
Under completion semantics, this report implicitly claims the existence of a completion witness satisfying the conditions of \autoref{def:comp-witness}.
Consequently, a natural form of self-validation is to {\em prompt} the same model to generate a completion witness explicitly.
The key intuition is that: if the LLM can correctly reason that a property holds under some completion, then it should also be capable of constructing a completion exhibiting the property.
Completion witness generation therefore converts an implicit inference into an explicit artifact that can be inspected and validated.

More concretely, completion witness generation
decomposes the original inference into three separate obligations to address the concerns identified in \autoref{sec:approach}:
\begin{enumerate}[leftmargin=*]
\item {\bf Completion construction.}
Generate a completion that makes explicit the assumptions required for the inference.
\item {\bf Domain validation.}
Establish that the completion is within the semantic domain of the query.
\item {\bf Property validation.}
Establish that the queried property holds for the completion under the underlying semantics and underlying reasoning procedure.
\end{enumerate}
For LLM-induced completion models, the first obligation is naturally performed by the LLM itself as a form of {\em self-validation}.
That is, the model is challenged to construct the omitted program context that explains or justifies the reported inference, and the resulting completion serves as a concrete realization of assumptions that were previously implicit.
The second and third properties can be established using {\em external oracles}.
Depending on the underlying semantics, these oracles may consist of compilers, interpreters, theorem provers, symbolic executors, testing frameworks, or other program-analysis tools.
Completion witness generation itself is therefore not tied to a particular semantic domain; rather, it provides a general mechanism for connecting completion semantics with existing validation procedures.

\subsection{Execution Witnesses}

Completion semantics is agnostic to the underlying semantic relation ($\models$) used to evaluate queries, meaning that the witness definition of \autoref{sec:witness-def} therefore supports arbitrary semantic domains in principle.
In this section, we focus on {\em execution-based} queries of the form:
\begin{align}
\prog \models \exists e : q(e)
\tag{\sc Existential-Execution}
\end{align}
where $q(e)$ is predicate over an {\em execution} $e$ of program $P$.
Here, an {\em execution} is any {\em semantic object}  that can be used to represent the concept of an execution through program $P$, which is typically defined by the programming language semantics.
As with the definition of ($\models$) itself, we also deliberately keep the definition of {\em execution} abstract, as there are many possible ways to represent an execution through a program (e.g., {\em traces} and {\em inputs}).
An {\em execution witness} is therefore a concrete execution $w$ such that $q(w)$ holds.

For example, for any valid completion $C$ of \autoref{fig:curl-strcpy}, an {\em execution witness} would be any execution that induces an overflow in \verb+strcpy+.
The execution witness may be represented as an input to the completion witness $W$ that induces the overflow.
The overflow can be detected by a reasoning procedure, which can be a bug detection oracle such as AddressSanitizer~\cite{konstantin2012asan}.

\subsubsection*{Universal Queries}
As with completion witnesses, an execution witness for the negated query can be used to {\em refute} universal queries of the form: $\prog \models \forall e : q(e)$.

\subsection{Witness Generation Workflow}

The witness-generation process combines the completion-witness and execution-witness concepts introduced in the previous sections. Given a fragment $\frag$, instruction $\instr$, and an existential inference:
\begin{align}
\Mllm(\frag) \models_{\exists} \exists e : q(e)
\tag{\sc existential-query}
\end{align}
the goal is to construct a witness pair
$\langle W, w \rangle$
consisting of a {\em completion witness} $W$ and an {\em execution witness} $w$.
The completion required for validation is typically much smaller than a fully functional program.
Rather, witness generation may seek a {\em minimal completion} sufficient to explain the reported inference.
In practice, such completions often resemble a test {\em harness} that supplies the minimal context required for the property to hold.

Conceptually, the workflow proceeds in two stages.
First, the LLM is tasked with {\em generating} a completion witness $W$ that makes explicit the assumptions underlying the reported inference.
The LLM is explicitly instructed to find a minimal completion $W$ such that:
\begin{enumerate}[leftmargin=*]
\item \label{case:member} $W \in \Mllm(\frag)$;
\item \label{case:dom} $W$ compiles and runs (i.e., $W$ lies within the semantic domain of the query); and
\item \label{case:exec} $W$ accepts an input that induces an execution $w$ such that $q(w)$ holds.
\end{enumerate}
Since membership in $\Mllm(\frag)$ has no independent oracle, completion generation must necessarily rely on the LLM itself (i.e., self-validation).
However, both (\ref{case:dom}) and (\ref{case:exec}) can be validated with external oracles.
As with completion generation, execution-witness generation may be performed iteratively, using feedback from the oracle to refine candidate witnesses.

\subsubsection*{Example}
When applied to the vulnerability report of \autoref{fig:curl-strcpy}, the workflow attempts to construct both a completion explaining the reported overflow and an execution demonstrating the overflow.
Failure to generate such a witness pair provides evidence that the reported vulnerability is unsupported by the completion model explored by the recovery procedure.

\subsection{Implementation}
\begin{figure}[t]
    \centering
    \includegraphics[width=.9\linewidth]{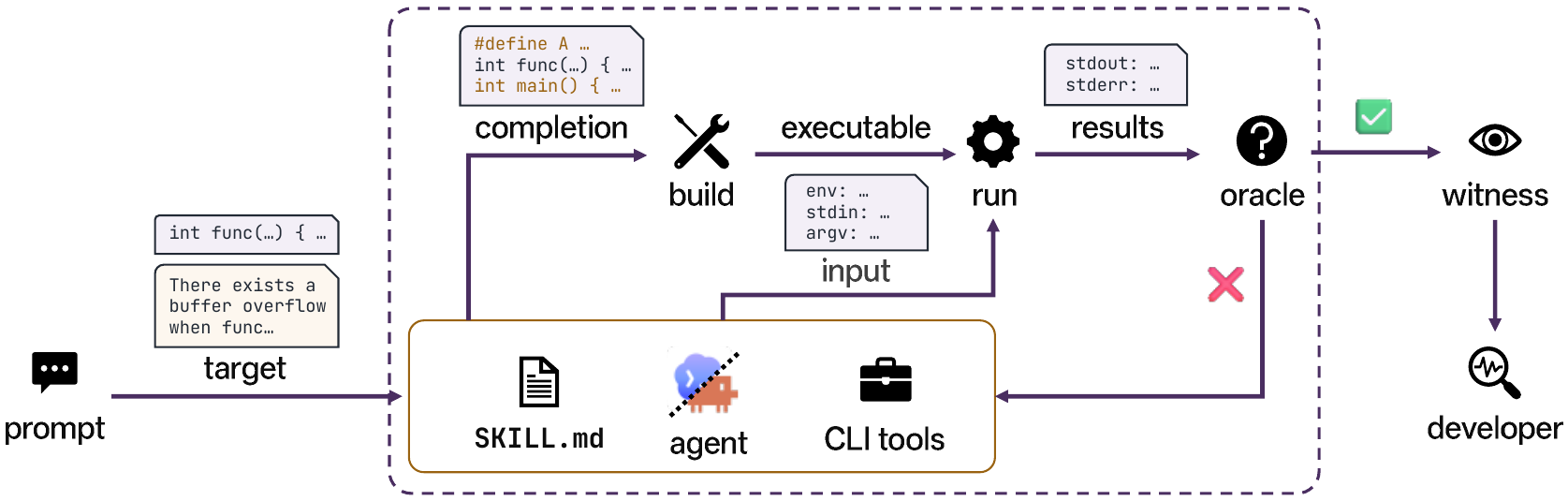}
    \caption{The \tool workflow.}
    \label{fig:workflow}
\end{figure}

We have implemented the workflow as an AI coding agent {\em skill}~\cite{anthropic2026skills} that is compatible with modern coding agents, such as Codex~\cite{openai2026codex} and Claude Code~\cite{anthropic2026claudecode}.
Here, an agentic {\em skill} is a modular bundle of instructions, resources, and executable components that agents can dynamically invoke for specialized tasks.
\tool is our implementation of a witness generation agent skill.

An overview of the \tool is shown in \autoref{fig:workflow}.
The core orchestration is handled by the agent, and the core functionality is handled by a \verb|witgen| CLI tool, implemented in Python, that supports candidate completion witness generation, building, and running.
Leveraging the CLI tool, \tool works in a refinement loop:
given a \textit{target} consisting of a code fragment and an existential claim, \tool first completes the code fragment under the assumptions that explain the inference.
Next, \tool attempts to build and run the completion with synthesized inputs and specified environments, such as command-line arguments and environment variables.
Then, the execution results are judged by an external oracle other than \tool itself, like embedded assertions, sanitizers, \etc
If a candidate is generated, but the oracle fails to confirm the claimed property (\eg no buffer overflow detected), the agent will refine the completion or propose a new one in the next iteration; otherwise, if this successfully yields a valid completion and execution witness that proves the claim, or the agent reaches a preset maximal retry count, then the process is over with a reported verdict: supported witness or no witness found.

Additionally, by adjusting the instructions in the \textit{skill}, the agent can be guided to generate completions following different disciplines, such as preserving standard library behavior or rewriting it, based on the LLM's knowledge and understanding from training data and its reasoning capabilities.
This unlocks \tool to generate witnesses according to different completion models, \eg compilation models and contract models mentioned in \autoref{sec:model-def}, enabling us to reveal witnesses in different contexts during evaluation.

\section{Evaluation}\label{sec:evaluation}

\subsection{Experimental Setup}

\myparagraph{Dataset Selection} To evaluate \tool and completion semantics, we use several datasets, including both {\em true positives} (TP) where the inference should hold, and {\em false positives} (FP) where the inference should not hold.
The TP/FP judgment is made by an external oracle.
The datasets are:

\begin{itemize}[leftmargin=*]
\item \textsc{CVEs}: 50 real-world CVEs from \textsc{MegaVul}~\cite{MegaVul}.
All subjects are TPs.
\item \textsc{AI-Slop}: 12 false bug reports (all FPs) listed under the \verb+cURL+ project's ``{\em AI slop hall of shame}''\footnote{See \url{https://gist.github.com/bagder/07f7581f6e3d78ef37dfbfc81fd1d1cd?permalink_comment_id=5956460} (retrieved July 2026).}.
We treat the \verb+cURL+ developer's rejection of these bug reports as the external oracle.
\item \textsc{AutoBug}: 30 subjects from the \textsc{Mixed-Manual} dataset from \textsc{AutoBug}~\cite{AutoBug}.
Out of the 30 subjects, 15 are reported as TPs, and the other 15 are reported as FPs.
We treat the original test harness for these subjects as the external oracle.
\end{itemize}
The dataset selection provides a balance between valid report, and false reports as judged by an external oracle.
For the \textsc{AI-Slop}, we selected the 12 usable reports out of 49 candidates that (1) specifically identified a bug, and (2) specifically identified a source location.
The remaining 38 reports were not bugs (e.g., missing documentation), reported alleged buggy behavior but failed to identify a location, or appeared to be entirely hallucinated (reported location does not exist).
Both \textsc{CVEs} and \textsc{AI-Slop} focus on existential inference in the form of bug reports, whereas \textsc{AutoBug} is existential inference in the form of post-condition violations represented by an \verb+assert+ statement.

\begin{table}
\centering
\small
\caption{Summary of results in \tool evaluation.
Here, \emph{Found} is the number of witnesses found, \emph{Unrealistic} is the number of {\em unrealistic} witnesses found (as deemed by an external oracle), {\em Not Found} is the number of subjects where no witness was found, and {\em Aligned} is how closely the result aligns with expectations (found for TP, not found for FP).
\label{tab:witgen-results}}
\begin{tabular}{lcrcccc}
\toprule
\textbf{Dataset} &
\textbf{Subset} &
\textbf{Total} &
\textbf{Found} &
\textbf{Unrealistic} &
\textbf{Not Found} &
\textbf{Aligned} \\
\hline
\hline

\textsc{CVEs}
& TP & 50 & 43 (86.0\%) & 0 (0.0\%) & 7 (14.0\%) & {\bf 43 (86.0\%)} \\
\hline

\textsc{AI-Slop}
& FP & 12 & 0 (0.0\%) & 4 (33.3\%) & 8 (66.7\%) & {\bf 8 (66.7\%)} \\
\hline

\multirow{2}{*}{\textsc{AutoBug}}
& TP & 15 & 15 (100.0\%) & 0 (0.0\%) & 0 (0.0\%) & {\bf 15 (100.0\%)} \\
& FP & 15 & 0 (0.0\%) & 7 (46.7\%) & 8 (53.3\%) & {\bf 8 (53.3\%)} \\
\bottomrule

\end{tabular}
\end{table}

\myparagraph{Dataset Configuration} For each test subject in our dataset, we construct natural-language instructions (e.g., existence of a {\em bug} or {\em assertion failure}) with a corresponding code fragment (relevant function).
For the \textsc{CVEs} dataset, we extract the instruction from the {\em Common Vulnerabilities and Exposures} (CVE) description and the original issue report, and the affected function from the corresponding \textsc{MegaVul} metadata.
For the \textsc{AI-Slop} dataset, we manually select reports that refer to concrete code within the \verb+cURL+ codebase, retrieve the relevant code fragment, and condense each report into concise one- or two-sentence instructions.
Finally, for \textsc{AutoBug}, we used the original subjects explicitly marked pre- and post-conditions.
We then formulate a claim that the post-condition assertion fails for some input satisfying the pre-condition.
These subjects exhibit a range of behaviors.
The \textsc{CVEs} dataset tests real-world vulnerabilities to test the effectiveness of \tool against {\em true positives} (TPs) as independently determined by a third party.
The \textsc{AI-Slop} tests \tool against {\em false positives} (FPs) as determined by a third party (the \verb+cURL+ developers).
Finally, the \textsc{AutoBug} dataset tests \tool against \verb+C+ and Python code subjects obtained from solutions to coding contest problems.
Here, a subject is considered to be a TP if there exists an input that satisfies the pre-condition, and yet violates the post-condition.
In the evaluation, we use the GPT-5.4~\cite{GPT-5.4} frontier model as the underlying LLM in \tool's workflow.
We use Codex~\cite{openai2026codex} as the underlying agentic architecture.
The \tool loop is run for a maximum of 3 iterations.

\myparagraph{Prompt (Re)Construction}
For some subjects, the original prompt and LLM (if any) used to make the inference were never made public and are deemed lost.
Thus, for these experiments, we run \tool using a {\em reconstructed prompt} comprising a natural language {\em description} of the inference (bug type and location) as the instruction $\instr$, and the function where the bug occurs as the code fragment $\frag$.
We run \tool under the assumption that the prompt had resulted in the corresponding inference, and \tool attempts to generate a witness accordingly.

\myparagraph{Variants}
We also test three main variants of our setup:
\begin{enumerate}[leftmargin=*]
\item \label{case:default} \emph{Default}: Attempts to find a completion consistent with $\Mllm(\frag)$;
\item \label{case:hack} \emph{+Hack}: Attempts to find {\em any} completion that compiles and triggers the claim; and
\item \label{case:context} \emph{+Context}: Same as (\ref{case:default}), but manually adds additional context to the prompt to further constrain $\Mllm(\frag)$ to match developer intent.
\end{enumerate}
Here, (\ref{case:default}) is our main result.
The (\ref{case:hack}) variant uses a modified version of \tool that instructs the LLM to generate a witness under $\M_\textsc{All}$---including any witness that would generally be considered to be unreasonable.
The (\ref{case:context}) variant is used to distinguish prompting errors from other hallucinations.

\begin{table}
\centering
\scriptsize
\caption{Table Of Results In \tool Evaluation On \textsc{AI-Slop} and \textsc{AutoBug} false positives (FP).
Here ``{\em Not Found}'' means that no witness was found (the expected result for a FP) and ``{\em Unrealistic}'' means a witness was found, but deemed unrealistic by the external oracle.\label{tab:witgen-results-detailed}}
\begin{tabular}{c|c|cc|cc|cc}
\toprule
&
\textbf{Id} &
\textbf{+Hack} &
\textbf{Explanation} &
\textbf{Default} &
\textbf{Explanation} &
\textbf{+Context} &
\textbf{Explanation}
\\ \hline
\hline

\multirow{12}{*}{\rotatebox[origin=c]{90}{\textsc{AI-Slop}}}
& \verb|2298307| & \cellcolor{yellow!20}Unrealistic & Mismatched \verb|payload| & \cellcolor{green!20}\cmark Not Found & & \cellcolor{green!20}\cmark Not Found & = \\
& \verb|2823554| & \cellcolor{yellow!20}Unrealistic & Injected \verb|strlen| & \cellcolor{green!20}\cmark Not Found & & \cellcolor{green!20}\cmark Not Found & = \\
& \verb|2887487| & \cellcolor{yellow!20}Unrealistic & Mismatched \verb|size| & \cellcolor{green!20}\cmark Not Found & & \cellcolor{green!20}\cmark Not Found & = \\
& \verb|2981245| & \cellcolor{yellow!20}Unrealistic & Dangling parameter & \cellcolor{green!20}\cmark Not Found & & \cellcolor{green!20}\cmark Not Found & = \\
& \verb|3117697| & \cellcolor{yellow!20}Unrealistic & Circular reference & \cellcolor{yellow!20}Unrealistic & = & \cellcolor{green!20}\cmark Not Found & \makecell{+ \texttt{Cookie} invariant} \\
& \verb|3137657| & \cellcolor{green!20}\cmark Not Found & & \cellcolor{green!20}\cmark Not Found & = & \cellcolor{green!20}\cmark Not Found & = \\
& \verb|3230082| & \cellcolor{green!20}\cmark Not Found & & \cellcolor{green!20}\cmark Not Found & = & \cellcolor{green!20}\cmark Not Found & = \\
& \verb|3392174| & \cellcolor{yellow!20}Unrealistic & Mismatched \verb|outlen| & \cellcolor{green!20}\cmark Not Found & & \cellcolor{green!20}\cmark Not Found & = \\
& \verb|3459636| & \cellcolor{yellow!20}Unrealistic & Injected \verb|strcpy| & \cellcolor{green!20}\cmark Not Found & & \cellcolor{green!20}\cmark Not Found & = \\
& \verb|3462525| & \cellcolor{yellow!20}Unrealistic & \autoref{sec:case-study-1} & \cellcolor{yellow!20}Unrealistic & = & \cellcolor{green!20}\cmark Not Found & + \verb|buf| space pre-cond. \\
& \verb|3516186| & \cellcolor{yellow!20}Unrealistic & \autoref{sec:case-study-2} & \cellcolor{yellow!20}Unrealistic & = & \cellcolor{green!20}\cmark Not Found & + \verb|time_t| non-negative \\
& \verb|3516202| & \cellcolor{yellow!20}Unrealistic & \autoref{sec:case-study-3} & \cellcolor{yellow!20}Unrealistic & = & \cellcolor{green!20}\cmark Not Found & + Single threaded \\

\hline

\multirow{15}{*}{\rotatebox[origin=c]{90}{\textsc{AutoBug} (FP)}}
& \verb|q0003| & \cellcolor{yellow!20}Unrealistic & Invalid UTF-8 & \cellcolor{yellow!20}Unrealistic & = & \cellcolor{green!20}\cmark Not Found & + \verb|s.isascii()| \\
& \verb|q0007| & \cellcolor{green!20}\cmark Not Found & & \cellcolor{green!20}\cmark Not Found & = & \cellcolor{green!20}\cmark Not Found & = \\
& \verb|q0069| & \cellcolor{green!20}\cmark Not Found & & \cellcolor{green!20}\cmark Not Found & = & \cellcolor{green!20}\cmark Not Found & = \\
& \verb|q0080| & \cellcolor{green!20}\cmark Not Found & & \cellcolor{green!20}\cmark Not Found & = & \cellcolor{green!20}\cmark Not Found & = \\
& \verb|q0161| & \cellcolor{green!20}\cmark Not Found & & \cellcolor{green!20}\cmark Not Found & = & \cellcolor{green!20}\cmark Not Found & = \\
& \verb|q0238| & \cellcolor{yellow!20}Unrealistic & Integer overflow & \cellcolor{yellow!20}Unrealistic & = & \cellcolor{green!20}\cmark Not Found & + Element range \\
& \verb|task23| & \cellcolor{green!20}\cmark Not Found & & \cellcolor{green!20}\cmark Not Found & = & \cellcolor{green!20}\cmark Not Found & = \\
& \verb|task36| & \cellcolor{green!20}\cmark Not Found &  & \cellcolor{green!20}\cmark Not Found & = & \cellcolor{green!20}\cmark Not Found & = \\
& \verb|task51| & \cellcolor{yellow!20}Unrealistic & \verb|NaN| parameter & \cellcolor{yellow!20}Unrealistic & = & \cellcolor{green!20}\cmark Not Found & + Element range \\
& \verb|task54| & \cellcolor{yellow!20}Unrealistic & \verb|NaN| parameter & \cellcolor{green!20}\cmark Not Found & & \cellcolor{green!20}\cmark Not Found & = \\
& \verb|task58| & \cellcolor{yellow!20}Unrealistic & Missing \verb|year| limits & \cellcolor{yellow!20}Unrealistic & = & \cellcolor{green!20}\cmark Not Found & + Element range \\
& \verb|task61| & \cellcolor{yellow!20}Unrealistic & Dynamic typing &  \cellcolor{yellow!20}Unrealistic & = & \cellcolor{green!20}\cmark Not Found & + Type hint \\
& \verb|task69| & \cellcolor{yellow!20}Unrealistic & Dynamic typing &  \cellcolor{yellow!20}Unrealistic & = & \cellcolor{green!20}\cmark Not Found & + Type hint \\
& \verb|task72| & \cellcolor{yellow!20}Unrealistic & Dynamic typing & \cellcolor{green!20}\cmark Not Found & & \cellcolor{green!20}\cmark Not Found & = \\
& \verb|task84| & \cellcolor{yellow!20}Unrealistic & UTF-8 \verb|swapcase()| & \cellcolor{yellow!20}Unrealistic & Dynamic typing & \cellcolor{green!20}\cmark Not Found & + \verb|s.isascii()| \\

\bottomrule

\end{tabular}
\end{table}

\subsection{Insights from Witness Generation}

The objective of our evaluation is to investigate whether witness generation can expose assumptions predicted by the completion semantics foundation.
Consequently, successful witness generation provides evidence that the reported inference is supported under the explored completion model, while failure to recover such a witness may indicate either a recovery limitation or a mismatch between the assumptions required by the inference and those admitted by the completion model.
A summary of the overall results is shown in \autoref{tab:witgen-results}, and detailed results for the FP subjects are showing in \autoref{tab:witgen-results-detailed}.
The main findings are discussed below.

\myparagraph{True Positives Admit Witnesses}
Across the \textsc{MegaVul} and \textsc{AutoBug} (TP) datasets, \tool consistently generates executable witnesses for the majority of subjects (\autoref{tab:witgen-results}).
This result is consistent with the completion semantics interpretation of existential reasoning: if a reported vulnerability is {\em realizable} under the completion model explored by the LLM, then there should exist a completion witness together with a corresponding execution witness demonstrating the property.
Cases where no witness was recovered tend to correspond to practical limitations of witness recovery (recovery errors) rather than evidence that the underlying vulnerability does not exist.

\myparagraph{False Positives Tend to Lack Admissible Witnesses}
The \textsc{AI-Slop} and \textsc{AutoBug} (FP) datasets may initially appear {\em plausible} when considered over incomplete program fragments alone.
However, when \tool attempts to explicitly construct a completion witness, the process often fails, or underlying (unreasonable) assumptions tend to become exposed.
For the majority of reports (66.7\% for \textsc{AI-Slop} and 53.3\% for \textsc{AutoBug} (FP)), \tool fails to generate {\em any} reasonable completion.
In other cases, the workflow is able to construct witnesses, but only by introducing assumptions that were deemed implausible or inconsistent with the surrounding code, especially for \autoref{tab:witgen-results-detailed} +Hack mode, where unreasonable witnesses are expressly allowed.
These examples support the central claim of this paper: the disagreement in inference results is not simply LLM ``{\em hallucination}''.
Rather, the reported inference depends on a completion model, which can differ from that implicitly assumed by the developers.
We will discuss some case studies in more detail in \autoref{sec:case-study}.

\myparagraph{Additional Context Resolves Completion Model Mismatch}
To further investigate unrealistic witnesses, we performed an additional +Context experiment in which the original prompt was manually extended with additional context or pre-conditions that mirror the external oracle.
Recall that we use {\em systematic} prompt (re)construction for Default: the prompt fragment $\frag$ contains only the relevant function and no other context, which is sometimes insufficient.
The +Context experiment tests a prediction made by completion semantics: that unrealistic witnesses can arise from overly permissive completion models that arise from under-constrained prompts.
Our results support this hypothesis.
For most unrealistic witnesses, additional context can be provided to eliminate unrealistic completions.
This suggests that many unrealistic witnesses should not necessarily be interpreted as failures of the LLM's reasoning procedure or of witness generation itself, but rather as consequences of an under-constrained prompt inducing an overly permissive completion model.

\myparagraph{Discussion}
Overall, our evaluation supports the completion semantics perspective developed throughout this paper.
Witness generation is valuable not merely because it validates individual LLM-generated inferences, but because it makes explicit the assumptions required for those inferences to hold.
When witnesses can be constructed, they provide concrete evidence supporting an existential claim.
When witness generation fails, or only succeeds under unrealistic assumptions, the failure itself is informative: it often reveals a mismatch between competing completion models rather than a simple reasoning error.
We believe this ability to expose hidden assumptions is at least as valuable as the witnesses themselves, since it provides developers with an explanation of \emph{why} different analyses may legitimately disagree over incomplete programs.

\subsection{Case Studies}\label{sec:case-study}

We consider three case studies from the \textsc{AI-Slop} dataset that highlight the challenges with LLM-based program inference.
Despite these bug reports being rejected by the \verb+cURL+ developers, many reports nevertheless support interpretations where the inference is {\em correct}.
We discuss below.

\subsubsection{Case Study: Missing Calling Assumptions}\label{sec:case-study-1}
Our first case study\footnote{\url{https://hackerone.com/reports/3462525}} illustrates the simplest form of completion ambiguity.
The report concerns the following small helper function:

\newpage

{\small 
\begin{Verbatim}[commandchars=\\\{\}, frame=single]
    \textcolor{darkgreen}{static int} storebuffer(\textcolor{darkgreen}{unsigned char} outc, \textcolor{darkgreen}{void} *f) \{
      \textcolor{darkgreen}{char} **buffer = f;
      **buffer = (\textcolor{darkgreen}{char})outc;       \textcolor{darkmagenta}{// Store outc into the buffer}
      (*buffer)++;                 \textcolor{darkmagenta}{// Increment the buffer pointer}
      \textcolor{darkyellow}{return} \textcolor{darkred}{0};
    \}
\end{Verbatim}
}
A call to \verb+storebuffer(ch, &buf)+ essentially just implements the operation \verb_*buf++ = ch_.
When presented with this function in isolation, most LLMs will flag a potential buffer overflow hazard, since the function does no explicit bounds checking internally.
Furthermore, we can use \tool to synthesize a witness program that (1) allocates a single-byte buffer, and (2) invokes \verb+storebuffer(...)+ twice to cause an overflow in the second call.
The overflow is readily detected using AddressSanitizer.
The inference is thus {\bf correct} under the LLM's completion model ($\Mllm$).

The \verb+cURL+ developers nevertheless rejected the report.
Their explanation was not that the synthesized witness is necessarily incorrect, but rather that it violates an implicit assumption of the function's API.
Specifically, \verb+storebuffer(...)+ is only meant to be called when the {\em caller} has verified that sufficient buffer space remains.
Consequently, the witness generated by the LLM does not reflect any intended or actual usage of the function.
The inference is therefore {\bf incorrect} under the developer's implicit completion model.

This case study illustrates an example of {\em prompting failure}.
Specifically, the prompt omitted critical information necessary to constrain the completion model to match that of the developer, allowing the LLM to reason over a substantially larger space of admissible completions than intended.
Importantly, the disagreement did not arise due to either party performing incorrect reasoning.
Rather, both the LLM and the developers derive valid conclusions with respect to their respective completion models.
This simple example also highlights how prompt construction for LLM-based program reasoning is non-trivial in the general case.

\subsubsection{Case Study: Single-Threaded versus Multi-Threaded Execution Model}\label{sec:case-study-3}

Our second case study\footnote{\url{https://hackerone.com/reports/3516202}} alleges a use-after-free vulnerability in the \verb+replace_existing(...)+ function from the \verb+cURL+ cookie subsystem.
The report claims that ``{\em the function modifies a linked list while iterating over it}'', creating the potential for memory corruption in concurrent environments.
The report assumes that multiple invocations of the function may execute concurrently on the same cookie data structure.
Under such a completion model, the inference is {\bf correct}, since one thread may retain a pointer to a linked-list node while another thread concurrently removes and frees the same node.
Since the function performs no synchronization, such an execution may result in a use-after-free.

However, the \verb+cURL+ developers promptly rejected the report, explaining that the function is never executed concurrently and that ``{\em each \verb+cURL+ invoke runs isolated from the others}'' so ``{\em there isn't really a race condition}''.
Under a completion model that only admits {\em single-threaded} execution environments (consistent with the actual \verb+cURL+ code base), access to the cookie database is externally serialized.
Consequently, the reported execution cannot occur, and the inference is {\bf incorrect}.

This example illustrates how completion models not only describe missing source code, but also constrain the execution environment in which the fragment executes.
Here, the relevant distinction is not the implementation of any omitted function, but whether the surrounding context permits concurrent execution.
The same program fragment, therefore, admits different conclusions depending solely on which execution model is considered admissible.

\subsubsection{Case Study: Integer Width Assumptions}\label{sec:case-study-2}

Another report\footnote{\url{https://hackerone.com/reports/3516186}} submitted to the \verb+cURL+ bug bounty program alleged an integer clamping bug when processing the \verb+Max-Age+ attribute of HTTP cookies:

{\small 
\begin{Verbatim}[commandchars=\\\{\}, frame=single]
  \textcolor{darkgreen}{time_t} now = time(\textcolor{darkred}{NULL});
  \textcolor{darkyellow}{if} (\textcolor{darkred}{CURL_OFF_T_MAX} - now < co->expires)   \textcolor{darkmagenta}{// Does expires+now exceeds the max value?}
    co->expires = \textcolor{darkred}{CURL_OFF_T_MAX};           \textcolor{darkmagenta}{// If yes, then clamp}
  \textcolor{darkyellow}{else}
    co->expires += now;                     \textcolor{darkmagenta}{// If no, then accumulate normally}
\end{Verbatim}
}
This code ensures the sum \verb_co->expires+now_ is representable as a 64-bit signed integer, or else the result is clamped to the max value (\verb+CURL_OFF_T_MAX+).
However, the bug report alleges that when ``{\em the current time (\verb+now+) is large enough}'', then the subtraction ``{\em produces unexpected results}'', allowing for the \texttt{else}-branch to be taken in cases where the sum can overflow or wrap.

The report was promptly dismissed by the \verb+cURL+ developers, who noted that
``{\em for this to be true, surely \verb+now+ would have to be larger than \verb+CURL_OFF_T_MAX+}?''
Under the actual \verb+cURL+ implementation, an implicit representation assumption ($\texttt{CURL\_OFF\_T\_MAX} \geq \texttt{now}$) will always hold, since \verb+CURL_OFF_T_MAX+ is the maximum possible value (\verb+INT64_MAX+) and that \verb+now+/\verb+expires+ are non-negative values.
This means the reported bug is impossible in practice, and any inference under such a model will be {\bf incorrect}.
The \autoref{tab:witgen-results-detailed} +Context result makes these assumptions explicit.

However, this inference assumes that the \verb+time_t+ type is a 64-bit unsigned integer.
While this is true for all modern operating systems, the \verb+C+ standard does not technically guarantee this.
Instead, the \verb+C+ standard merely requires \verb+time_t+ to be an integer type of some unspecified signedness and width.
Under this specification, the \verb+C+ standard would technically allow a {\em 65-bit width} for \verb+time_t+:

{\small
\begin{Verbatim}[commandchars=\\\{\}]
    \textcolor{darkyellow}{typedef} \textcolor{darkgreen}{unsigned} \textcolor{darkgreen}{_BitInt}(\textcolor{darkred}{65}) \textcolor{darkgreen}{time_t};
\end{Verbatim}
}
Such a completion would allow for values that directly violate the invariant, for example:

{\small
\begin{Verbatim}[commandchars=\\\{\}]
    now = \textcolor{darkred}{CURL_OFF_T_MAX} + \textcolor{darkred}{1}
\end{Verbatim}
}
Under such a completion model, the \verb+if+-conditional is {\em not} sufficient to prevent the \texttt{else}-branch from being taken when wrapping could occur,
meaning the inference is \textbf{correct}.

This example illustrates an important feature of completion semantics.
The disagreement is not fundamentally about arithmetic, but about omitted assumptions.
Under the intended \verb+cURL+ completion model with the ubiquitous \verb+time_t+ definition, the report is invalid.
However, under a more permissive completion model admitting all \verb+C+-conforming implementations, the same inference technically becomes valid.
The bug report therefore cannot be classified as simply ``correct'' or ``incorrect'' independently of the completion model against which it is interpreted.

\subsection{Empirical Study Comparison}\label{sec:empirical}

Recent empirical studies~\cite{nong2026llm, koterba2026llm, yang2025llm, li2025everything, fang2024llm} have collectively identified a consistent set of phenomena regarding LLM-based program analysis, particularly for {\em vulnerability detection}, {\em classification}, and {\em repair}.
Across different models, benchmarks, and prompting strategies, these studies repeatedly report that the quality of LLM-based program reasoning heavily depends on contextual information, assumptions about omitted code, and prior exposure to similar implementations.
Collectively, these observations suggest that many apparent failures of LLM-based program analysis arise from reasoning over incomplete programs, rather than from reasoning errors alone.
A qualitative analysis mapping these empirical works to completion semantics is shown in \autoref{tab:empirical}.
In this section, we summarize the major findings and relate them back to completion semantics.

\myparagraph{Context is Important}
Many empirical studies~\cite{nong2026llm, koterba2026llm, yang2025llm, li2025everything} identify {\em insufficient context} as a major factor in the quality of LLM-based code inference (\autoref{tab:empirical}~(1), (2), and (8)).
For example, especially with {\em vulnerability detection}, most studies found that {\em function-level} analysis is generally insufficient, and advocate providing additional contextual information.
Completion semantics reaches a similar conclusion: insufficient context leaves the completion model under-constrained, allowing for the admission of reasonable program completions that disagree with the developer's expectations or the code base.
This is reflected in our \autoref{tab:witgen-results-detailed} Default and +Context results, where the difference in alignment is explained by missing context.

Most existing studies conclude that more context {\em should} be included into prompts.
However, the recent work of~\cite{yang2025llm} also demonstrates that indiscriminately supplying additional context can also {\em reduce} analysis accuracy due to the introduction of ``excessive noise''.
This indicates that the objective is not to maximize the amount of code presented to the LLM, but to provide minimal information that most effectively constrains the intended completion model to the intended target.
Consequently, practical LLM-based analysis involves selecting a context that best approximates the intended semantic completion while remaining within the reasoning capabilities of the model.

\myparagraph{Many Factors Shape LLM Completion Models}
A second recurring observation is that a wide variety of seemingly unrelated factors influence LLM reasoning (\autoref{tab:empirical} (3)-(6)), including assumptions about omitted APIs, identifier naming, code obfuscation, and prior exposure to similar code during training.
From the perspective of completion semantics, these factors all influence the shape of the induced completion model ($\Mllm$) by changing which program completions are considered admissible.
Our framework provides a common semantic interpretation for these otherwise disparate empirical observations.
At the same time, completion semantics is largely agnostic as to how strongly each factor influences the induced completion model.
This is illustrated by \tool---which self-validates against the LLM's own completion model---independent of empirical observations.

\myparagraph{Completion Models need not Contain Well-formed Programs}
Finally, it has been observed that LLMs may generate code that does not compile~\cite{fang2024llm}.
This motivates one of our design choices: completion models are not restricted to syntactically or semantically well-formed programs.
Instead, they represent the space of candidate completions that could be considered during inference, while independent oracles (e.g., compilers and test suites) may subsequently reject invalid completions.

\myparagraph{Discussion}
Our work is complementary to these empirical studies.
Rather than evaluating prompting strategies or proposing new prompts, we ask a more fundamental semantic question: {\em what does it mean for an LLM-generated inference over an incomplete program fragment to be correct?}
We argue that incomplete code fragments do not denote a single program, but rather a space of possible completions, and that correctness must therefore be interpreted relative to a completion model.
From this perspective, many reported ``false positives'' are not necessarily incorrect inferences, but instead reflect different assumptions about omitted code, contracts, and execution context.
Consequently, our framework complements empirical observations by explaining why these limitations arise as an inherent consequence of reasoning over incomplete programs.

\begin{table}[t]
\centering
\scriptsize
\caption{Examples of prior empirical observations and their interpretation under completion semantics.}
\label{tab:empirical}
\begin{tabular}{p{0.2cm}p{6.5cm}p{5.8cm}}
\toprule
\textbf{No.}
&
\textbf{Empirical observation}
&
\textbf{Completion semantics interpretation}
\\
\midrule

\cellcolor{black!5}{(1)}
&
\cellcolor{black!5}{\makecell[l]{LLM conclusions change as additional code context is provided. \\ \cite{koterba2026llm, li2025everything, nong2026llm, yang2025llm}}}
&
\vspace{-1.3em}
\cellcolor{black!5}{Additional context rules out previously admissible completions.}
\\[1ex]

\cellcolor{black!0}{(2)}
&
\cellcolor{black!0}{Redundant code context can degrade LLM analysis accuracy~\cite{yang2025llm}.}
&
\cellcolor{black!0}{The representation supplied to the LLM is no longer an efficient encoding of the intended completion model.}
\\

\cellcolor{black!5}{(3)}
&
\cellcolor{black!5}{\makecell[l]{LLMs make incorrect assumptions about missing code or APIs. \\ \cite{li2025everything, nong2026llm, yang2025llm}}}
&
\vspace{-1.3em}
\cellcolor{black!5}{The LLM reasons under a completion model that differs from the developer's intended completion model or the original code base.}
\\[1ex]

\cellcolor{black!0}{(4)}
&
\cellcolor{black!0}{Meaningful identifier names improve analysis accuracy~\cite{fang2024llm}.}
&
\cellcolor{black!0}{Identifier names constrain the completion model by excluding
otherwise plausible interpretations.}
\\[1ex]

\cellcolor{black!5}{(5)}
&
\cellcolor{black!5}{Code obfuscation reduces analysis accuracy~\cite{fang2024llm, koterba2026llm}.}
&
\cellcolor{black!5}{Obfuscation removes semantic constraints from the fragment,
expanding the completion model and increasing ambiguity.}
\\[1ex]

\cellcolor{black!0}{(6)}
&
\cellcolor{black!0}{\makecell[l]{LLM analyses are influenced by prior exposure to similar code. \\
\cite{fang2024llm, koterba2026llm, nong2026llm}}}
&
\vspace{-1.3em}
\cellcolor{black!0}{Previously observed implementations (during training) can bias the completion model
towards particular completions.}
\\[1ex]

\cellcolor{black!5}{(7)}
&
\cellcolor{black!5}{LLMs often generate broken code that does not compile~\cite{fang2024llm}.}
&
\cellcolor{black!5}{The completion model does not necessarily only contain compilable-programs.}
\\[1ex]

\cellcolor{black!0}{(8)}
&
\cellcolor{black!0}{\makecell[l]{LLM analyses can produce false positives and false negatives. \\ \cite{koterba2026llm, li2025everything, nong2026llm, yang2025llm}}}
&
\vspace{-1.3em}
\cellcolor{black!0}{These may arise either from reasoning errors under the chosen
completion model or from mismatch between the induced completion model and the intended developer model.}
\\

\bottomrule
\end{tabular}
\end{table}

\section{Related Work}

\myparagraph{Classical Program Semantics}
Classical program semantic frameworks~\cite{Floyd67, Dijkstra75, Hoare69} generally define reasoning over complete programs by first interpreting programs within some {\em semantic domain} (e.g., {\em traces}, {\em transition systems}, {\em abstract domains}, or {\em logical formulae}).
Completion semantics instead addresses an orthogonal question: {\em how should incomplete program fragments be interpreted before such semantics are applied?}
Rather than replacing existing semantic frameworks, completion semantics lifts them to incomplete programs by interpreting a fragment as a set of admissible completions.
Any underlying program semantics can then be applied unchanged to each completion.
Likewise, completion semantics is independent of the reasoning procedure used to establish properties over those semantics.
This separation is particularly natural for LLM-based program analysis, where the reasoning procedure is an {\em approximate} language model operating directly over the source text.

\myparagraph{Modular Program Reasoning}
Classical approaches to modular program reasoning~\cite{meyer1997object, pierce2002types, chandy1988parallel}, including {\em contract-based verification}, {\em assume-guarantee reasoning}, and {\em type systems}, also reason about incomplete programs.
Rather than requiring a complete program, these techniques use {\em interfaces}, {\em contracts}, or other {\em specifications} to characterize the behavior of omitted components.
Such specifications effectively define a completion model by restricting which implementations of the missing components are considered admissible.
Classical modular reasoning typically assumes specifications are explicitly provided and can be reasoned about formally.
In contrast, LLM-based reasoning can be understood in terms of a completion model built from the prompt, surrounding code, and prior knowledge encoded by the LLM.

\myparagraph{Specification Inference}
Related work on {\em specification inference}~\cite{ernst2007daikon, flanagan2001houdini}, including LLM-based specification inference~\cite{ma2025specgen, endres2024spec, chapman2024llm}, attempts to {\em recover} assumptions about missing program behaviour automatically.
Such inferred specifications can be understood as constraints over, or descriptions of, possible completion models.
Completion semantics is intentionally more general, as admissible completions need not be expressible as behavioural specifications alone.
Missing {\em type definitions}, {\em data structure layouts}, {\em libraries}, or {\em environmental assumptions} may all influence the meaning of an incomplete fragment without naturally corresponding to traditional specifications.
Completion semantics therefore separates the question of which completions are admissible from the particular formalism used to describe them, allowing both explicit specifications and implicitly inferred assumptions to be treated within a common semantic framework.

\myparagraph{Program Slicing}
Program slicing~\cite{korel1988slice, weiser1981slice} seeks to identify the {\em subset} of a program relevant to a computation or property.
\textsc{AutoBug}~\cite{AutoBug} and \textsc{HyllFuzz}~\cite{meng2026hyllfuzz} are examples of slice-based LLM program reasoning systems. 
In terms of completion semantics, the slice may correspond to an ``optimized'' fragment whose omitted context ought not alter the semantics.
Slicing and completion semantics address different questions: slicing attempts to {\em minimize context}, whereas completion semantics assigns meaning to the ultimate fragment.
Slicing may still recursively introduce dependencies as relevant procedures, types, and global state, which can still be of significant size.
Thus, even slice-based systems tend to omit context, meaning that completion semantics remains applicable.

\myparagraph{LLM-based Program Reasoning}
Beyond empirical studies of \autoref{sec:empirical}, several works~\cite{becker26llm, chen2025llm, gu2024crux, li2024pbe} have investigated the broader problem of {\em LLM-based program reasoning}.
Becker et al.~\cite{becker26llm} study reasoning over source code as a program-analysis task, characterizing the capabilities and limitations of LLMs when performing semantic reasoning about programs.
$\mathcal{R}$Eval~\cite{chen2025llm} extends this direction by introducing benchmarks that evaluate reasoning about program runtime behaviour (execution paths and program state), together with a notion of incremental reasoning consistency.
Crucially, these works primarily investigate how well LLMs reason about {\em complete executable programs} that have a limited size.
In contrast, our work addresses a complementary semantic question: because LLMs over large code bases will typically reason over incomplete program fragments, the correctness of an inference necessarily depends on assumptions about omitted context.

\myparagraph{LLMs and Artifact Generation}
Recent work~\cite{zhao2026anypoc, ahmed2025otter, yang2025verus, yang2026exverus} has increasingly coupled LLM reasoning with the generation of {\em artifacts} that can be independently validated.
Examples include executable proof-of-concept exploits and regression tests for validating bug reports (e.g., AnyPoC~\cite{zhao2026anypoc} and Otter~\cite{ahmed2025otter}), as well as machine-checkable proofs, verification annotations, and counterexamples in LLM-assisted formal verification systems such as AutoVerus~\cite{yang2025verus} and ExVerus~\cite{yang2026exverus}.
Our work is complementary to this direction.
We do not claim artifact generation as a contribution in itself; rather, we provide a semantic foundation explaining why such artifacts are meaningful for reasoning over incomplete programs.

\myparagraph{AI Coding Agents}
Recent coding agents such as Claude Code~\cite{anthropic2026claudecode} and Codex~\cite{openai2026codex} mitigate limited context by iteratively retrieving additional files, invoking search tools, and requesting clarifications when insufficient information is available.
This changes the practical workflow, but not the underlying semantic problem considered in this paper.
At every stage, the agent reasons over a finite context while treating the remainder of the codebase and execution environment as implicit.
Even when an agent decides that additional context is required, it must eventually terminate retrieval and perform inference relative to the remaining omitted context.
Completion semantics therefore applies equally to agentic workflows: the retrieval strategy simply changes where the boundary between explicit and implicit context is drawn.

\section{Discussion}

The primary contribution of this paper is a new semantic framework for program reasoning over incomplete programs.
{\em Completion semantics} is intentionally straightforward: an incomplete program fragment effectively denotes a space of possible completions, and correctness is defined relative to this space.
This formalism leads to several useful observations regarding the nature of LLM-based program reasoning that are difficult to express without first making completion models explicit.
We discuss the main findings in this section.

\myparagraph{Correctness is Relative}
The motivating question of this paper is: \emph{when is an LLM-based program reasoning correct?}
Completion semantics suggests that this question does not admit an absolute answer.
Since LLMs are typically forced to reason over incomplete program fragments, and fragments do not denote a single program, then correctness is necessarily {\em relative} to the completion model used to interpret the fragment.

This perspective explains why apparently contradictory conclusions may simultaneously be reasonable.
The examples from \autoref{fig:example} and \autoref{fig:curl-strcpy}, and the case studies from \autoref{sec:case-study}, all illustrate this phenomenon.
The same program fragment will frequently admit both ``correct'' and ``incorrect'' interpretations depending on the assumptions made that are encoded in the completion model, specifically regarding omitted {\em types}, {\em functions}, {\em contracts}, {\em execution environments}, or {\em platform-specific behavior}.
Likewise, the case study in \autoref{sec:case-study-2} demonstrates how even seemingly objective notions, such as ``{\em the \verb+C+ language semantics}'', can lead to surprising inferences.
Disagreements therefore do not imply incorrect reasoning; they may instead arise because different parties reason under different completion models.
Completion semantics therefore shifts the question from ``{\em is the inference correct?}'' to ``{\em under which completion model is the inference correct?}''

\myparagraph{Context Selection Problem}
Completion semantics also highlight a fundamental tension in LLM-based program reasoning.
Adding additional context to a prompt generally improves the accuracy, with respect to developer expectations, of the induced completion model.
However, larger prompts also consume finite reasoning resources, increase retrieval complexity, and may reduce reasoning quality due to context dilution or LLM cognitive overload.
Conversely, aggressively minimizing prompts may improve reasoning efficiency, but it enlarges the completion space, increasing the likelihood that the model reasons under assumptions that differ from those intended.

This trade-off appears to be inherent rather than incidental.
Modern software systems are sufficiently large that an LLM cannot reason over an entire codebase and its execution environment simultaneously.
Regardless of how much context is supplied, a boundary between observed and omitted context will usually exist.
Prompt construction therefore becomes an optimization problem balancing completion-model fidelity against available reasoning resources.
Completion semantics describes this trade-off independently of any particular retrieval strategy or language model.

\myparagraph{Reasoning Failures are Not Monolithic}
Current discussions of LLM reliability often describe incorrect outputs simply as ``{\em hallucinations}.'' Completion semantics suggests that this view is overly coarse.
\autoref{sec:failure} identifies several distinct failure modes.
An inference may fail because the LLM's completion model itself differs from the developer's intended interpretation.
Alternatively, the completion model may be appropriate, while the reasoning procedure itself reaches the wrong conclusion.
Further failures arise when attempting to recover explicit witnesses from an implicit completion model.
These failure modes have different causes and therefore should have different remedies.
Completion-model mismatch motivates improved retrieval or prompt construction.
Reasoning and recovery failures motivate improved (frontier) models.
Reasoning errors also motivate validation methods such as witness generation.
Distinguishing these cases provides a more precise understanding of LLM behavior.

\myparagraph{Implications for Evaluation}
Completion semantics also have implications for the evaluation of LLM-based program reasoning systems.
Many existing benchmarks classify generated reports simply as correct or incorrect.
Completion semantics suggests that such labels implicitly assume a particular completion model, even if that model is never explicitly stated.
Consequently, benchmark disagreements may reflect differences in assumed context rather than deficiencies of the underlying reasoning procedure.
This observation suggests that future evaluations should distinguish reasoning quality from completion-model quality whenever possible.
For example, witness generation allows an inferred completion to be inspected directly, making it possible to determine whether a disagreement arises because the reasoning procedure failed or because the assumptions underlying the inference differ from those intended by the developer.

\myparagraph{Broader Perspective}
Completion semantics are intentionally independent of any particular reasoning engine.
Throughout this paper, we focused primarily on LLM-based reasoning as a natural application of completion models.
However, any reasoning over incomplete programs may be interpreted through the same completion-model perspective.
Likewise, witness generation should be viewed as one possible instantiation of the framework (rather than the framework itself).
Witnesses provide a practical mechanism for exposing assumptions underlying existential inferences, but other mechanisms for exploring or characterizing completion models are possible.

\subsection*{Acknowledgements}

This research is supported by the National Research Foundation, Singapore, under its National Cybersecurity R\&D Programme (Award No. CRPO-GC5-NUS-004).

\bibliographystyle{ACM-Reference-Format}
\bibliography{main}

\end{document}